\documentclass[12pt,a4paper]{article}
\pdfoutput=1

\usepackage{jheplike,ifpdf,tocloft,rotating,amsmath,amssymb,mathrsfs,extarrows,mathtools,subcaption,hyperref,microtype,color,xcolor,multirow}
\usepackage{float}

\widowpenalty=500\clubpenalty=1000\unitlength=1mm\textheight=8.7in\hypersetup{pdftitle={},pdfcreator={},linkcolor=[rgb]{0.15,0.35,0.75},colorlinks=true,citecolor=[rgb]{0.675,0,0.2},urlcolor=[rgb]{0.15,0.35,0.65}}
\let\olditemize\itemize\renewcommand{\itemize}{\vspace{-2pt}\olditemize\setlength{\itemsep}{1pt}\setlength{\parskip}{0pt}\setlength{\parsep}{-0pt}}
\let\oldenumerate\enumerate\renewcommand{\enumerate}{\vspace{-4pt}\oldenumerate\setlength{\itemsep}{1pt}\setlength{\parskip}{0pt}\setlength{\parsep}{0pt}}
\makeatletter\renewcommand\section{\addtocontents{toc}{\protect\addvspace{-2.25\p@}}\@startsection {section}{1}{\z@}{0.5ex \@plus .2ex \@minus 0.2ex}{0.3ex \@plus.1ex\@minus .5ex}{\normalfont\large\bfseries}}
\renewcommand\subsection{\addtocontents{toc}{\protect\addvspace{0.5\p@}}\@startsection {subsection}{1}{\z@}{0.5ex \@plus .2ex \@minus 0.2ex}{0.3ex \@plus.1ex\@minus .5ex}{\normalfont\bfseries}}
\renewcommand\subsubsection{\addtocontents{toc}{\protect\addvspace{-2.5\p@}}\@startsection {subsubsection}{1}{\z@}{0.5ex \@plus .2ex \@minus 0.2ex}{0.3ex \@plus.1ex\@minus .5ex}{\normalfont\bfseries}}
\counterwithout{paragraph}{subsubsection}

\newcommand{\eq}[1]{\vspace{-0.5pt}\begin{equation}#1\vspace{-0.5pt}\end{equation}}

\newcommand{\pl}{\raisebox{0.75pt}{\scalebox{0.75}{$\hspace{-2pt}\,+\,\hspace{-0.5pt}$}}}

\renewcommand{\tilde}{\widetilde}
\renewcommand{\phi}{\varphi}

\newcommand{\x}[2]{(#1,#2)}

\newcommand{\proj}[1]{\raisebox{1pt}{{\big[}}#1\raisebox{1pt}{{\big]}}}

\newcommand{\traintrack}{\mathfrak{T}^{(L)}}

\definecolor{hblue}{rgb}{0,0,0.575}
\definecolor{hred}{rgb}{0.575,0.0,0.225}
\definecolor{hgreen}{rgb}{0.0,0.4,0.2}
\definecolor{dim}{rgb}{0.55,0.55,0.55}

\newcommand{\ud}[2]{\genfrac{}{}{0pt}0{#1}{#2}}

\title{Cutting the traintracks: Cauchy, Schubert and Calabi-Yau}
\author[a,f]{\vspace{-24pt}Qu Cao,}\emailAdd{qucao@zju.edu.cn}
\author[a,b,c,d]{Song He,}\emailAdd{songhe@itp.ac.cn}
\author[a,e]{Yichao Tang}\emailAdd{tangyichao@itp.ac.cn}
\affiliation[a]{CAS Key Laboratory of Theoretical Physics, Institute of Theoretical Physics, Chinese Academy of Sciences, Beijing 100190, China}
\affiliation[b]{School of Fundamental Physics and Mathematical Sciences, Hangzhou Institute for Advanced Study}
\affiliation[c]{International Centre for Theoretical Physics Asia-Pacific, Beijing/Hangzhou, China}
\affiliation[d]{Peng Huanwu Center for Fundamental Theory, Hefei, Anhui 230026, P. R. China}
\affiliation[e]{School of Physical Sciences, University of Chinese Academy of Sciences, No.19A Yuquan Road, Beijing 100049, China}
\affiliation[f]{Zhejiang Institute of Modern Physics, Department of Physics, Zhejiang University, Hangzhou, 310027, China
}

\abstract{
In this note we revisit the maximal-codimension residues, or leading singularities, of four-dimensional $L$-loop traintrack integrals with massive legs, both in Feynman parameter space and in momentum (twistor) space. We identify a class of ``half traintracks" as the most general degenerations of traintracks with conventional (0-form) leading singularities, although the integrals themselves still have rigidity $\lfloor\frac{L-1}2\rfloor$ due to lower-loop ``full traintrack'' subtopologies. As a warm-up exercise, we derive closed-form expressions for their leading singularities both via (Cauchy's) residues in Feynman parameters, and more geometrically using the so-called Schubert problems in momentum twistor space. For $L$-loop full traintracks, we compute their leading singularities as integrals of $(L{-}1)$-forms, which proves that the rigidity is $L{-}1$ as expected; the form is given by an inverse square root of an irreducible polynomial quartic with respect to each variable, which characterizes an $(L{-}1)$-dim Calabi-Yau manifold (elliptic curve, K3 surface, {\it etc.}) for any $L$.
We also briefly comment on the implications for the ``symbology" of these traintrack integrals. }
\preprint{}

\begin{document}
\maketitle\thispagestyle{empty}


\vspace{\baselineskip}
\section{Introduction and summary of main results}
Multi-loop Feynman integrals are both central objects and major bottlenecks for our ability to make precise predictions in Quantum Field Theory. In the last decades, enormous progress has been made in Feynman integrals as well as physical quantities such as scattering amplitudes which evaluate to the simplest class of functions, multiple polylogarithms (MPL)~\cite{Chen:1977oja,Goncharov1995GeometryOC,Goncharov:1998kja,Remiddi:1999ew,Borwein:1999js,Moch:2001zr}. In particular, a driving force in understanding the analytic properties and hidden mathematical structures is the powerful idea of symbol and coproducts of these functions~\cite{Goncharov2002GaloisSO,Goncharov:2010jf,Duhr:2011zq,Duhr:2012fh}. However, more complicated functions, such as elliptic generalizations of MPL (eMPL)~\cite{Laporta:2004rb,Muller-Stach:2012tgz,Brown2011MultipleEP,Bloch:2013tra,Adams:2013nia,Adams:2014vja,Adams:2015gva,Adams:2015ydq,Adams:2016xah,Adams:2017ejb,Adams:2017tga,Bogner:2017vim,Broedel:2017kkb,Broedel:2017siw,Adams:2018yfj,Broedel:2018iwv,Broedel:2018qkq,Honemann:2018mrb,Bogner:2019lfa,Broedel:2019hyg,Duhr:2019rrs,Walden:2020odh,Weinzierl:2020fyx,Giroux:2022wav,Kristensson:2021ani,Wilhelm:2022wow,Morales:2022csr}, also appear for Feynman integrals; see~\cite{Bourjaily:2022bwx} for a review, and references therein. 

\begin{figure}[H]
		\centering
	\includegraphics[width=0.9\textwidth]{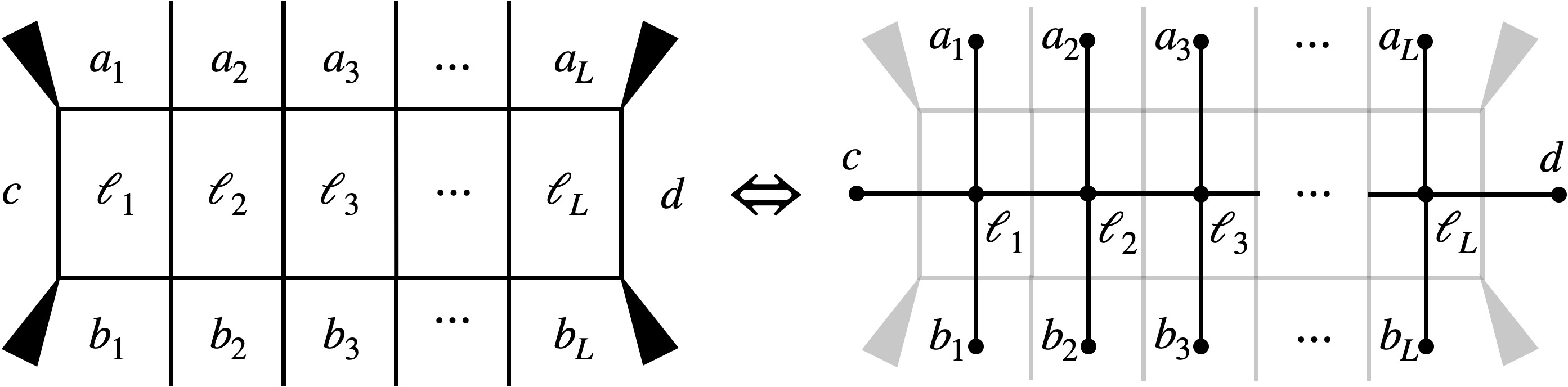}
 \caption{The massless traintrack integral and its dual graph}
	\label{fig:traintrack_massless_dual}
\end{figure}

In~\cite{Bourjaily:2018ycu}, the authors proposed to study a class of ``traintrack" integrals (Fig.~\ref{fig:traintrack_massless_dual}) as an interesting toy model because of the appearance of more and more complicated functions already for planar massless four-dimensional Feynman integrals. These integrals serve as an important ingredient for higher-order computations in a large class of theories such as $\lambda \phi^4$ and planar ${\cal N}=4$ super-Yang-Mills. It has been conjectured that an $L$-loop traintrack integral has ``rigidity" $L{-}1$ as it involves an $(L{-}1)$-fold integration over a certain Calabi-Yau (CY) manifold of dimension $L{-}1$; see~\cite{Bourjaily:2018yfy,Bourjaily:2022tep} for more detailed discussions on the concept of rigidity. For $L=1$, the box integral evaluates to MPL functions of rigidity 0, while for $L=2$, the famous double-box integral~\cite{Bourjaily:2017bsb} involves an elliptic curve and evaluates to eMPLs of rigidity 1\footnote{This fact has been known through the maximal cut and the Mellin representation in~\cite{Paulos:2012nu,Caron-Huot:2012awx,Nandan:2013ip,Chicherin:2017bxc}.}; the $L=3,4$ traintracks were argued to involve K3 surfaces and Calabi-Yau three-folds with rigidity 2 and 3 respectively in~\cite{Bourjaily:2018ycu, Bourjaily:2019hmc, Vergu:2020uur}.

An $L$-loop traintrack integral is defined in dual coordinate space as
\eq{\traintrack\!\equiv\!\int\!\!d^{4L}\!\vec{\ell}\frac{(c,d)\prod_{k=1}^L\x{a_k}{b_k}}{\x{\ell_1}{c}\Big[\prod_{k=1}^{L{-}1}\x{\ell_{k}}{\ell_{k+1}}\x{\ell_k}{a_k}\x{\ell_k}{b_k}\Big](\ell_L,a_L)(\ell_L,b_L)\x{\ell_L}{d}},\label{dual_space_integrand}}
where $(x,y)\equiv(x{-}y)^2$ denotes the squared distance between dual points, including the $L$ internal points $\{\ell_k\}_{k=1}^L$ dual to loop momenta and the $2L{+}2$ external ones.
The numerator ${\cal N}_L:=(c,d)\prod_{k=1}^L (a_k, b_k)$ has been introduced to ensure that the result is dual conformally invariant (DCI). We remind the readers that for a Feynman diagram in ordinary space, we can draw its dual graph (Fig.~\ref{fig:traintrack_massless_dual}), and the dual coordinate space is defined by expressing momenta as differences between dual points.

\begin{figure}[H]
		\centering
	\includegraphics[width=0.5\textwidth]{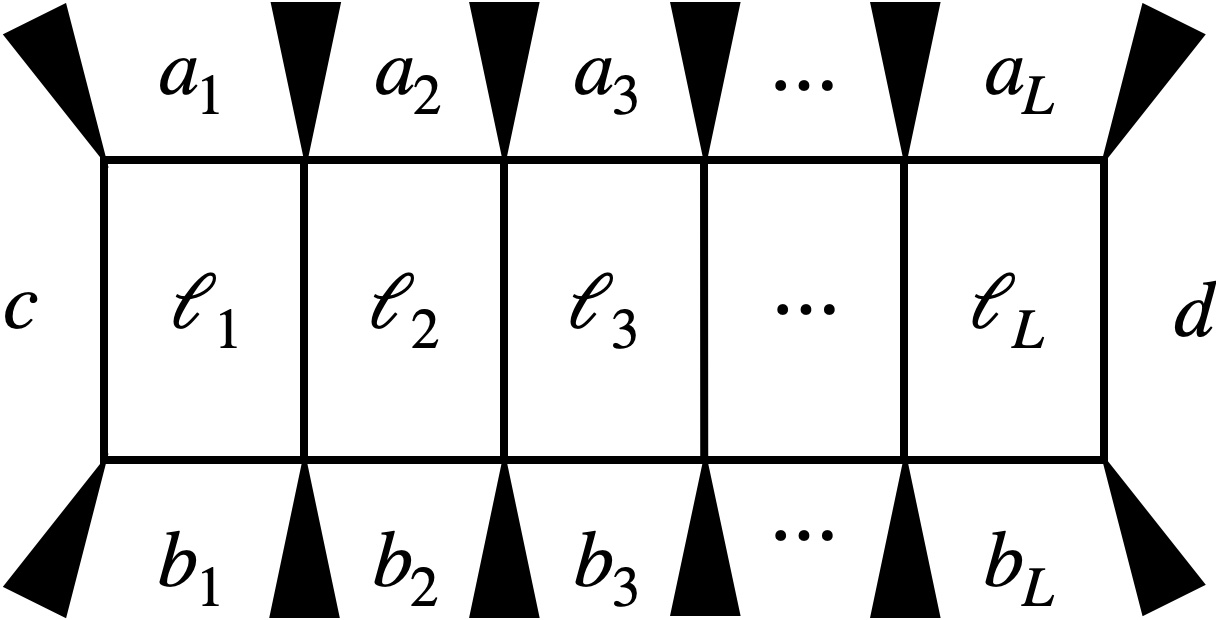}
	\caption{The massive full traintrack integral}
	\label{fig:traintrack_all}
\end{figure}

Originally, in~\cite{Bourjaily:2018ycu}, all $2(L{-}1)$ legs on the top and the bottom are taken to be massless, \emph{i.e.}, $(a_k, a_{k+1})=(b_k, b_{k+1})=0$ for $k=1, \cdots, L{-}1$; we refer to such an integral as a ``massless traintrack"\footnote{We always keep the four corners massive for simplicity, {\it i.e.}, $(a_1, c)$, $(b_1, c)$, $(a_L, d)$ and $(b_L,d)$ are all non-zero. The rigidity does drop if some corners become massless, but in that case, the integral suffers from infrared divergences and requires regularization.}. However, we drop these massless constraints and consider ``massive traintracks'' (Fig.~\ref{fig:traintrack_all}): this is not only the most general case, but they are also relevant even for massless traintracks since they appear as sub-topologies of massless ones. For $L>1$, it is straightforward to count that the massless traintrack depends on $6L{-}5$ independent DCI cross-ratios, while the massive traintrack depends on $8L{-}7$ cross-ratios\footnote{The well-known exception  at $L=1$ has 2 degrees of freedom instead of 1.}; the former corresponds to a $2(L{-}1)$-fold \emph{collinear} limit of the latter. 

Traintrack integrals have not been computed in general for $L>2$, and even for those degenerate cases that yield MPL functions, it gets more and more difficult to perform all integrations for larger $L$. A nice approach for analyzing the underlying mathematical structures of such integrals is through their ``leading singularities" (LS)~\cite{Bern:1994zx,Bern:1994cg,Cachazo:2008vp}, which can be computed by cutting all propagators\footnote{For LS computation of (non-DCI) integrals in dimensional regularization, see \emph{e.g.}, \cite{Frellesvig:2017aai,Bosma:2017ens,Harley:2017qut,Dlapa:2021qsl} using the Baikov representation, or \cite{Abreu:2017ptx,Flieger:2022xyq} in loop momentum space proper.} in loop momentum space~\cite{Arkani-Hamed:2010pyv,Caron-Huot:2012awx}, or similarly taking as many residues as possible in Feynman parameter space~\cite{Bourjaily:2018ycu,Bourjaily:2019hmc}. Recall that one sometimes have to take composite residues since in loop momentum space there are $3L{+}1$ propagators and $4L$ integrations, whereas with loop-by-loop Feynman parametrization, there are $L{+}1$ poles and $2L$ integrations. It has been conjectured in~\cite{Bourjaily:2018ycu} that, for ``full traintracks'' with $\{a_k,b_k\}_{k=1}^L$ all distinct, one arrives (in Feynman parameter space) at an $(L{-}1)$-form\footnote{One can compute periods by pairing such differential forms with certain choices of contours.
In this note we focus on the differential form (the ``integrand" of periods) and leave the study of integration contours to future work.} associated with a CY manifold, when no more residues can be taken of the integrand~\cite{Bourjaily:2020hjv}. However, to our best knowledge it has not been proven that the rigidity of the integral is indeed $L{-}1$, or that the resulting manifold must be CY for all $L$. Moreover, it is not at all obvious whether one would obtain the same LS by computing maximal-codimensional residues in Feynman parameter space and directly in loop momentum space; see~\cite{Frellesvig:2021vdl} for a non-trivial comparison of different parametrizations in elliptic cases.

In this note, we will perform a systematic all-loop study of the LS form for (massive) traintracks, and in particular we will prove that the rigidity of full traintracks is $L{-}1$ and that the resulting geometry is CY.

Before proceeding, let us already sketch the {\bf proof} of the rigidity with details to be filled in later. We begin with the loop-by-loop Feynman parametrization of an $L$-loop massive full traintrack. As we write explicitly in \eqref{eq-t-integral}, it can be written as an $(L{-}1)$-fold integral of a quadric integral, which is recognized as the Feynman-parametrized version of a deformed $(2L{+}2)$-gon integral and famously evaluates to weight-$(L{+}1)$ MPL functions~\cite{Spradlin:2011wp,Herrmann:2019upk,Bourjaily:2019exo,Arkani-Hamed:2017ahv}. Since the integral is expressed as an $(L{-}1)$-fold integral of MPL function,  we have $L{-}1$ as an {\it upper bound} of the rigidity. 

Next we prove that the upper bound is saturated, and we do so by explicitly computing the LS which amounts to changing the contour of integrations. It was known in~\cite{Bourjaily:2018ycu} that the LS computation provides a {\it lower bound} for the rigidity of integrals, which we will refer to as the {\bf rigidity of LS}; however, such computations have not been done correctly beyond $L=3$. As we show in section~\ref{sec-ls-cy}, after taking as many residues as possible, we are left with integrations over the remaining $L{-}1$ Feynman parameters, of the square root of an irreducible polynomial with degree four in each parameter. This implies that the rigidity $L{-}1$ is saturated. What is more, we will show in section~\ref{sec-ls-cy} that the polynomial defines an $(L{-}1)$-dim CY manifold independent of which $L{-}1$ variables are left to parametrize it! In fact, we will also perform an LS computation using momentum twistors parametrizing dual loop momenta (geometrically known as solving Schubert problems~\cite{Arkani-Hamed:2010pyv})
, which turns out to yield the square root of an irreducible polynomial of exactly the same degrees, thus also giving a CY manifold of dimension $L{-}1$.

\begin{table}[H]
	\centering
     \caption{\label{tab-comparsion} Basic countings for full and half traintrack integrals}
	\begin{tabular}{|c|c|c|c|c|}
		\hline
		&\multicolumn{2}{c|}{\multirow{2}{*}{full traintrack}}& \multicolumn{2}{c|}{\multirow{2}{*}{half traintrack}}\\
  & \multicolumn{2}{c|}{\quad}& \multicolumn{2}{c|}{\quad} \\
		\hline
		\multirow{2}{*}{number of external dual points} & \multicolumn{2}{c|}{\multirow{2}{*}{$2L+2$}} & \multicolumn{2}{c|}{\multirow{2}{*}{$L+3$}}\\
  & \multicolumn{2}{c|}{\quad}& \multicolumn{2}{c|}{\quad} \\
		\hline
		number of d.o.f. & massive & massless & massive & massless\\
		\cline{2-5}
		($L>1$)& $8L-7$ & $6L-5$ & $4L-3$&$3L-2$\\
		\hline rigidity of LS & \multicolumn{2}{c|}{\multirow{2}{*}{$L-1$}} & \multicolumn{2}{c|}{\multirow{2}{*}{$0$}}\\
		(dimension of LS form) & \multicolumn{2}{c|}{\quad}& \multicolumn{2}{c|}{\quad} \\
		\hline\multirow{2}{*}{rigidity (of integral)} & \multicolumn{2}{c|}{\multirow{2}{*}{$L-1$}} &  \multicolumn{2}{c|}{\multirow{2}{*}{$\displaystyle\left\lfloor\frac{L-1}{2}\right\rfloor$}}\\
  & \multicolumn{2}{c|}{\quad}& \multicolumn{2}{c|}{\quad} \\
		\hline
	\end{tabular}
\end{table}

Such LS computation is straightforward in principle but rather tedious in practice. As a warm-up exercise, we find it illuminating to first consider degenerations of traintrack integrals such that (composite) residues can be taken all the way down to give a $0$-form, as was originally meant by LS~\cite{Cachazo:2008vp}. This way, we find a zoo of interesting $L$-loop integrals with lower rigidity. Note that even though the LS is a $0$-form, there can still be lower-loop traintrack-type subtopologies (which we abbreviate as ``subtopologies'' in the following)\footnote{By this we mean a subgraph obtained by deleting or shrinking propagators, but which remains a traintrack integral. See Figs.~\ref{fig:integralA},~\ref{fig:integralB}, and~\ref{fig:integralC} for examples of this.} ,
which in turn put lower bounds on the rigidity. We will show that by taking $L{-}1$ {\it soft} limits, the most general traintracks with $0$-form LS are ``{\bf half traintracks}", where for each rung only one (possibly massive) leg remains either pointing upward or downward. In other words, we take either $a_k=a_{k+1}$ or $b_k=b_{k+1}$ but not both\footnote{We can take even more soft limits by letting $a_k=a_{k+1}$ and $b_k=b_{k+1}$ so that certain loops become ``ladder-like", but it is easy to see that the LS is unchanged when cutting these loops, a special case of the reduction process described in~\ref{ex-red-1} and~\ref{ex-red-2}. The most degenerate case is the ladder integral with all $a_k$ identified and all $b_k$ identified, which has been computed to all loops and even resummed~\cite{Usyukina:1993ch,Broadhurst:2010ds,Broadhurst:1993ib}; its LS is always the same as that of the one-loop box.} for $k=1,\cdots,L{-}1$, so that only $L{+}3$ distinct external points remain. Note that such an $L$-loop massive (resp., massless) half traintrack depends on $4L{-}3$ (resp., $3L{-}2$) DCI variables, roughly half the number in the full traintrack case. See Table~\ref{tab-comparsion} for a summary of these basic countings.

\begin{figure}[H]
\centering
\includegraphics[width=0.45\textwidth]{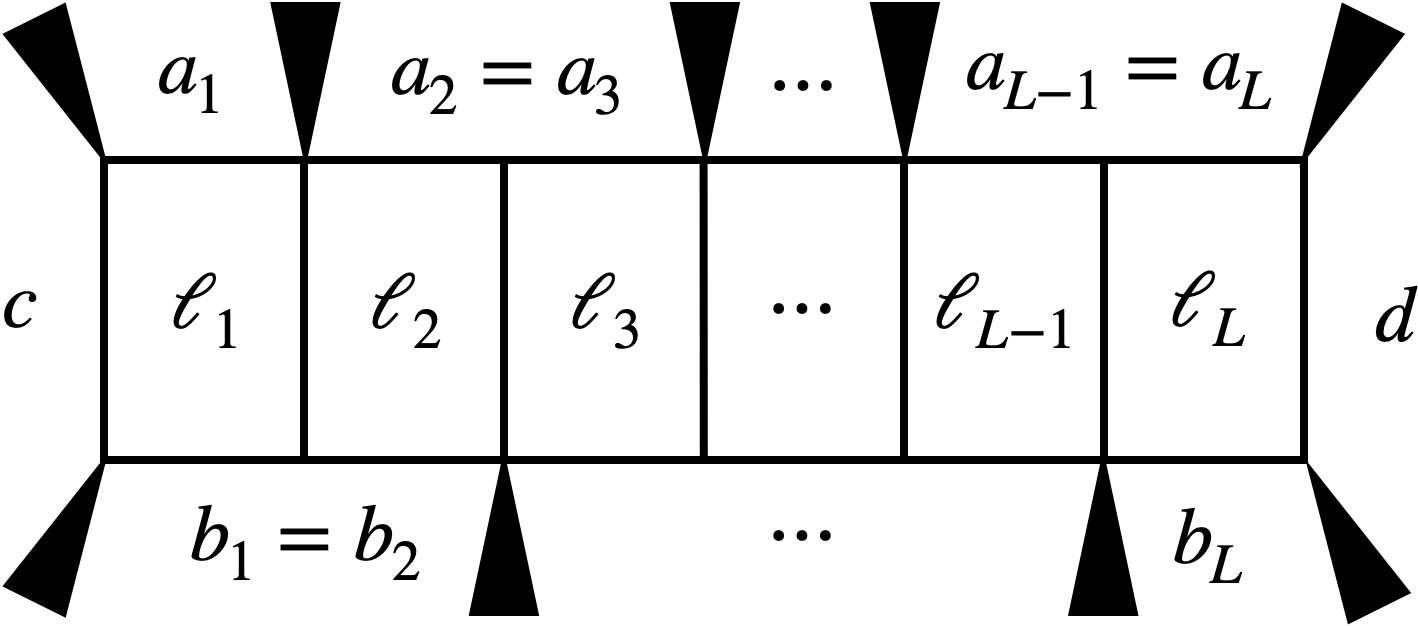}\qquad\includegraphics[width=0.45\textwidth]{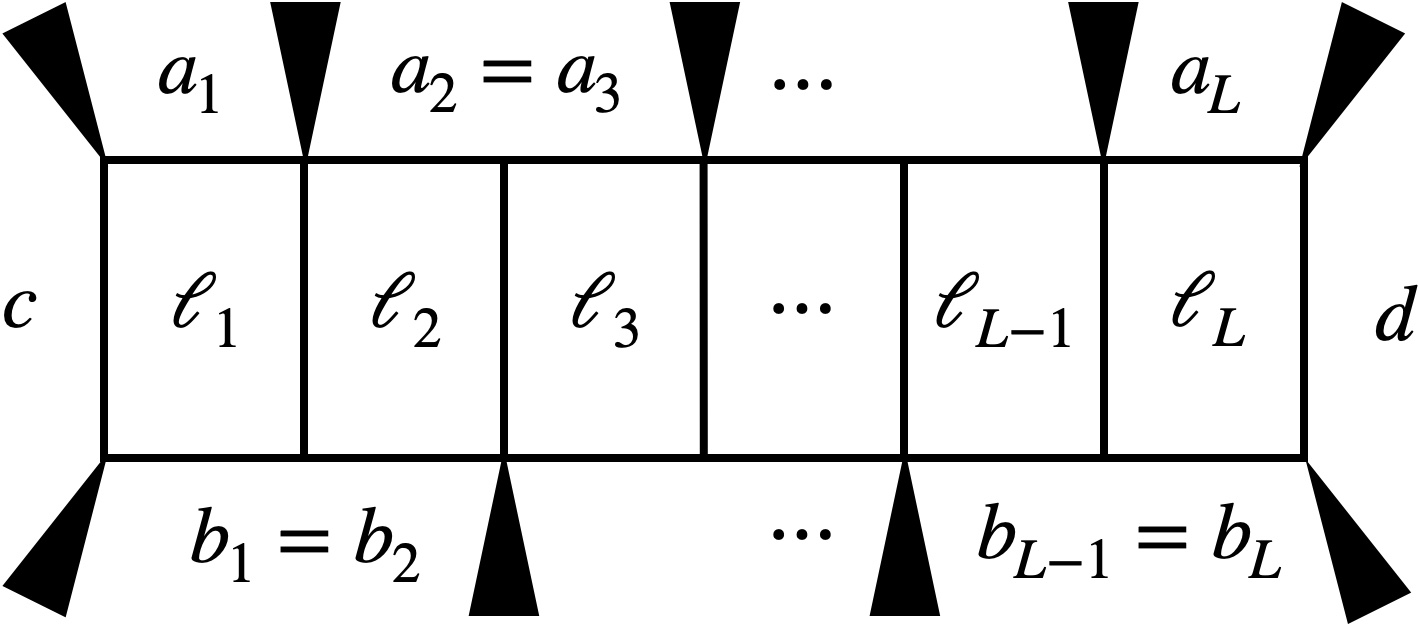}
	\caption{Alternating half traintrack integrals for odd and even $L$.}
	\label{fig:traintrack_even}
\end{figure}

Let us denote the full traintrack integral as well as its degenerations by the self-explanatory notation $\left[c\ud{a_1}{b_1}\ud{\cdots}{\cdots}\ud{a_L}{b_L}d\right]$. For example, for $L=2$ we have two possible half traintracks which are equivalent, $\left[c\ud{a_1}{b_1}\ud{a_1}{b_2}d\right]$ and $\left[c\ud{a_1}{b_1}\ud{a_2}{b_1}d\right]$. There are two inequivalent half traintracks for $L=3$: the ``all-up" traintrack $\left[c\ud{a_1}{b_1}\ud{a_2}{b_1}\ud{a_3}{b_1}d\right]$ and the ``alternating" traintrack $\left[c\ud{a_1}{b_1}\ud{a_2}{b_1}\ud{a_2}{b_3}d\right]$. For $L=4$, in addition to all-up and alternating cases, we have a third inequivalent half traintrack $\left[c\ud{a_1}{b_1}\ud{a_2}{b_1}\ud{a_3}{b_1}\ud{a_3}{b_4}d\right]$.
Our convention for the orientation of alternating traintracks is illustrated in Fig.~\ref{fig:traintrack_even}. It turns out that as far as LS of half traintracks are concerned, it suffices to consider the alternating cases whose LS depend on all external points, while the LS of other half traintracks effectively reduces to that of a lower-loop alternating one.

As the main result of section~\ref{sec-ls-0form}, we will derive the ($0$-form) LS for any half traintrack integral, both in Feynman parameter space and in momentum twistor space:
\begin{equation}
{\bf LS}^{(L)}_{\rm half}=\frac{{\cal N}_L}{\sqrt{\Delta_L}}\,,
\end{equation}
where ${\cal N}_L$ is the numerator in \eqref{dual_space_integrand} and $\Delta_L$ in the square root is the discriminant of a quadratic equation, with weight two in $a_k, b_k$ (for $k=1, \cdots, L$ which are identified in pairs), $c$ and $d$ such that the final result is DCI. 
Note that although half traintracks have $0$-form LS, generically the integrals themselves have non-zero rigidity since lower-loop subtopologies can have non-zero rigidity. For example, an $(R{+}1)$-loop full traintrack as subtopology raises the rigidity of a half traintrack to at least $R$. We will not provide the details here, but we have found that the most rigid subtopologies of any $L$-loop half traintrack are $\lfloor \frac{L{+}1}{2}\rfloor$-loop full traintrack integrals, regardless of the distribution of legs (all-up, alternating, or cases in between). Thus, the rigidity of any $L$-loop half traintrack is expected to be $\lfloor \frac{L{-}1}{2}\rfloor$ (see Table~\ref{tab-comparsion})!

The main result of section~\ref{sec-ls-cy} is that, for (massive) full traintrack integrals, in both Feynman parameter space and momentum (twistor) space, we are left with $(L{-}1)$-fold integrals when no more residues can be taken:
\begin{equation}
{\bf LS}^{(L)}_{\rm full}={\cal N}_L \int \frac{d^{L{-}1} \alpha}{\sqrt{{\cal Q}_L(\alpha})}\,,    
\end{equation}
where $\alpha:=\{\alpha_1,\cdots, \alpha_{L{-}1}\}$ denote the remaining $L{-}1$ variables (which have different meanings in the two spaces), and ${\cal Q}_L(\alpha)$ is an irreducible polynomial, which has weight two in each dual point such that the result is DCI. Remarkably we will see that our natural choices of taking residues lead to beautiful recursive formulas for $\mathcal Q_L(\alpha)$. We find that in both spaces, $\mathcal Q_L(\alpha)$ has degree 4 in each variable $\{\alpha_k\}_{k=1}^{L-1}$, which can be expressed as elliptic fibrations for {\it any} of the variables and defines a CY manifold embedded in $(\mathbb{CP}^{1})^L$. The two approaches lead to CY manifolds with similar defining polynomials. Classifications of CY manifolds of higher dimensions are difficult, so we have not been able to show that our two CY manifolds are isomorphic, although it is reasonable to suspect they are.

Another point worth mentioning is that these CY manifolds may develop singularities for large $L$, because the dimension of kinematic space grows linearly in $L$, whereas the dimension of moduli space of polynomials grows exponentially with $L$. In order to better study the intrinsic geometric properties, it might be helpful to find a CY embedding that is smooth for generic kinematics\footnote{This has been achieved for $L\leq4$ in~\cite{Vergu:2020uur}. However, it is not immediately clear how to generalize their method to higher loops.}. We leave this for future works.

\section{Leading singularities of half traintrack integrals}\label{sec-ls-0form}

In this section, we derive leading singularities of the most general half traintrack integrals first in Feynman parameter space and then in momentum (twistor) space, finding perfect agreement. We will see that an $L$-loop leading singularity is proportional to the inverse of $\sqrt{\Delta_L}$, where the ``discriminant" $\Delta_L=A_L^2- 4 B_L$ has weight two in each dual point before identification; in other words, if $a_k=a_{k+1}$, it would have a total weight of four in $\Delta_L$. Let us recall that for $L=1$, {\it i.e.}, a box integral $\left[c\ud{a}{b}d\right]$ with dual points $a,c,b,d$ in a cyclic ordering, we have
\begin{equation}\label{eq-A1}
A_1=(a,c)(b,d)+(a,d)(b,c)-(a,b)(c,d)\,,\quad
B_1=(a,c)(c,b)(b,d)(d,a)\,.
\end{equation}
Thus, with normalization factor ${\cal N}_1:=(a,b)(c,d)$, the LS reads 
\begin{equation}
{\bf LS}^{(1)}_\text{half}=\frac{(a,b)(c,d)}{\sqrt{A_1^2- 4  B_1}}=\frac{1}{\sqrt{(u+v-1)^2-4 u v}}\,,
\end{equation}
with DCI cross-ratios $u=\frac{(a,c)(b,d)}{(a,b)(c,d)}$ and $v=\frac{(a,d)(b,c)}{(a,b)(c,d)}$. Remarkably, all half traintrack LS turn out to be essentially ``one-loop-like": we find that $\Delta_L$ always has the form $A_L^2-4B_L$, where $A_L$ (resp., $B_L$) is a nice polynomial (resp., monomial) with the correct weights. Both of them have beautiful recursive structures; for the alternating case with $L\geq2$, $A_L$ is given by the determinant of an $(L{+}1)$-dim matrix $\mathcal A_L$ and it has $2^L$ terms, one of which is nothing but ${\cal N}_L$. As we shall soon see, the LS of general half traintracks reduce to those of alternating traintracks.

\subsection{Leading singularities in Feynman parameter space}\label{sec:fey-ls}

Let us begin with a loop-by-loop Feynman parametrization of massive traintracks, which are essentially the same as that of massless traintracks in~\cite{Bourjaily:2019jrk}:
\eq{\traintrack\!=\!\int_0^{\infty}\frac{{\rm d}^L\alpha\,{\rm d}^L\beta\,{\rm d}^L\gamma}{GL(1)^L}\frac{\mathcal{N}_L}{\x{R_1}{R_1}\cdots
\x{R_L}{R_L}\x{R_L}{d}},\label{inital_feynman_rep}}
where, within the embedding formalism,
\eq{\label{eq-gl1}
(R_k)\equiv\gamma_k(R_{k-1})+\alpha_k(a_k)+\beta_k(b_k), \quad\text{for }k=1,\cdots,L,\quad\text{and }R_0\equiv(c).}
In this section, for convenience, we will fix the $GL(1)$ redundancies by setting the Feynman parameters on the rungs to be $\gamma_k=1$.


For definiteness, we first consider the alternating cases (Fig.~\ref{fig:traintrack_even}) where $b_1=b_2$, $a_2=a_3$, ..., $b_{2k-1}=b_{2k}$, and $a_{2k}=a_{2k+1}$:
\begin{equation}
\traintrack_\text{alt}=\left[c\ud{a_1}{b_1}\ud{}{=}\ud{a_2}{b_2}\ud{=}{}\ud{\cdots}{\cdots}\ud{a_L}{b_L}d\right]=\int_0^\infty\frac{{\rm d}^L\alpha\,{\rm d}^L\beta\,\mathcal N_L}{(R_1,R_1)\cdots(R_L,R_L)(R_L,d)}.\label{inital_feynman_rep_alt}
\end{equation}


We evaluate the maximal-codimension (codim-$2L$) residue of eq.\eqref{inital_feynman_rep_alt} in a recursive manner: first, we take the residue at $(R_L,d)=(R_L,R_L)=0$ with respect to $\alpha_L$ and $\beta_L$ corresponding to the right-most loop; then, the ``half'' nature enables us to take the composite residue at $(R_k,R_k)=0$ with respect to $\alpha_k$ and $\beta_k$, iteratively for $k=L{-}1, \cdots, 1$.

Specifically, when we take the residue at $(R_L,d)=(R_L,R_L)=0$ with respect to $\alpha_L$ and $\beta_L$, we may presciently assume $(R_{L-1},R_{L-1})=0$ because the LS is supported on the vanishing set of $(R_{L-1},R_{L-1})$, among other factors. This way, the problem becomes that of finding the LS of the four-mass box $\left[R_{L-1}\ud{a_L}{b_L}d\right]$ with massless propagators. The result is $1/\sqrt{\Delta_{L}^{(1)}}$, where
\begin{equation}\label{eq-delta1}
	\begin{aligned}
		\Delta_{L}^{(1)}=&\left[\x{a_{L}}{d}\x{R_{L-1}}{b_{L}}+\x{b_{L}}{d}\x{R_{L-1}}{a_{L}}-\x{a_{L}}{b_{L}}\x{R_{L-1}}{d}\right]^2\\
		&-4\x{a_{L}}{d}\x{b_{L}}{d}\x{R_{L-1}}{a_{L}}\x{R_{L-1}}{b_{L}}\\
  \equiv&\left[A_{L}^{(1)}\right]^{2}-4B_{L}^{(1)}.
	\end{aligned}
\end{equation}
Here, the superscript $^{(1)}$ reminds us that this is only the intermediate result after step one; we have solved for $\alpha_L,\beta_L$ corresponding to the right-most loop. 

The next step is to take the composite residue at $(R_{L-1},R_{L-1})=0$ with respect to $\alpha_{L-1}$ and $\beta_{L-1}$. Again, we may assume $(R_{L-2},R_{L-2})=0$ on which the LS is supported. Then, recall that $\gamma_{L-1}=1$,
\begin{equation}
    \frac12(R_{L-1},R_{L-1})=\alpha_{L-1}(a_{L-1},R_{L-2})+\beta_{L-1}(b_{L-1},R_{L-2})+\alpha_{L-1}\beta_{L-1}(a_{L-1},b_{L-1}).
\end{equation}
There are many ways to compute the composite residue. In the alternating case, the most convenience way is to take the residue with respect to $\beta_{L-1}$ before $\alpha_{L-1}$ if $b_{L-1}=b_L$ (or in the opposite order if $a_{L-1}=a_L$). Specifically, if $b_{L-1}=b_{L}$, one could split eq.(\ref{eq-gl1}) into
\begin{equation}
    (R_{L-1})=(\tilde R_{L-1})+\beta_{L-1}(b_{L-1}),\quad(\tilde R_{L-1})=(R_{L-2})+\alpha_{L-1}(a_{L-1})\,,
\end{equation}
and take the residue at $\beta_{L-1}=-\frac{(\tilde R_{L-1},\tilde R_{L-1})}{2(b_{L-1},\tilde R_{L-1})}$ before using the Jacobian factor to take a further residue at $\alpha_{L-1}=-\frac{(b_{L-1},R_{L-2})}{(b_{L-1},a_{L-1})}$. Here, we only record the final result\footnote{We will ignore overall signs and factors of two.}:
\begin{equation}\label{eq-com-res}
    \mathop{\rm compRes}_{(R_{L-1},R_{L-1})=0}\frac1{(R_{L-1},R_{L-1})\sqrt{\Delta_L^{(1)}}}=\frac1{\sqrt{\Delta_L^{(2)}}},
\end{equation}
where
\begin{equation}
    \Delta_L^{(2)}=(a_{L-1},b_{L-1})^2\Delta_L^{(1)}\Big|_{\alpha_{L-1}\to-\frac{(b_{L-1},R_{L-2})}{(a_{L-1},b_{L-1})},\ \beta_{L-1}\to-\frac{(a_{L-1},R_{L-2})}{(a_{L-1},b_{L-1})}}\ .
\end{equation}
Here and after, we always use a vertical line to represent substitution/evaluation.
An equivalent description of the above rule of substitution is
\begin{equation}
    (R_{L-1},x)\to(R_{L-2},x)-\frac{(b_{L-1},R_{L-2})(a_{L-1},x)+(a_{L-1},R_{L-1})(b_{L-1},x)}{(a_{L-1},b_{L-1})}.
\end{equation}

For example, if $b_{L-1}=b_L$, we would find
\begin{align}
    \Delta_L^{(2)}&=\left[\x{b_{L}}{d}\det(a_{L-1},R_{L-2};a_{L},b_{L})-\x{a_{L}}{b_{L}}\det(a_{L-1},R_{L-2};d,b_{L})\right]^2\nonumber\\
    &-4\x{a_{L}}{d}\x{b_{L}}{d}\x{a_{L}}{b_{L}}\x{R_{L-2}}{a_{L-1}}\x{b_{L}}{R_{L-2}}\x{b_{L}}{a_{L-1}}\nonumber\\
    &\equiv\left[A_{L}^{(2)}\right]^{2}-4B_{L}^{(2)}.
\end{align}
Here, we have introduced the notation $\det(a,b;c,d)=\x{a}{c}\x{b}{d}-\x{a}{d}\x{b}{c}$. We remark that $A_L^{(2)}$ can be recognized as the determinant of a Gram matrix $\mathcal{A}_{L}^{(2)}$ with rows labelled by $\{b_{L},a_{L-1},R_{L-2}\}$ and columns labelled by $\{d,a_{L},b_{L}\}$:

 \begin{equation}\label{eq:amat2}
 \begin{aligned}
A_{L}^{(2)}=\det(\mathcal{A}_{L}^{(2)})&=\det\begin{pmatrix}
 		\x{d}{b_{L}}&\x{a_{L}}{b_{L}}\\ \det(a_{L-1},R_{L-2};d,b_{L})&\det(a_{L-1},R_{L-2};a_{L},b_{L})
\end{pmatrix}\\
&=\det\begin{pmatrix}
 		\x{d}{b_{L}}&\x{a_{L}}{b_{L}}&0\\ \x{d}{a_{L-1}}&\x{a_{L}}{a_{L-1}}&\x{b_{L}}{a_{L-1}}\\ \x{d}{R_{L-2}}&\x{a_{L}}{R_{L-2}}&\x{b_{L}}{R_{L-2}}
 		\end{pmatrix}.
 \end{aligned}
 \end{equation} 
The result is similar if $a_{L-1}=a_L$.

In general, by computing the composite residues at $(R_k,R_k)=0$ with respect to $\alpha_k$ and $\beta_k$ for $k=L-2,\cdots,1$, we are able to compute $\Delta_L^{(k)}=\left[A_L^{(k)}\right]^2-4B_L^{(k)}$ iteratively (more details can be found in Appendix~\ref{app-details-feynman}). The result is summarized by the following rule of substitution:
\begin{gather}
    (R_{L-k},x)\to(R_{L-k-1},x)-\frac{(b_{L-k},R_{L-k-1})(a_{L-k},x)+(a_{L-k},R_{L-k-1})(b_{L-k},x)}{(a_{L-k},b_{L-k})},\label{eq:subs-half}\\
    \Delta_L^{(k+1)}=(a_{L-k},b_{L-k})^2\Delta_L^{(k)}\Big|_{\text{eq.(\ref{eq:subs-half})}}\ .
\end{gather}
Equivalently,
\begin{gather}
    A_{L}^{(k+1)}=A_L^{(k)}\left\{\begin{aligned}
&(R_{L-k},x)\to \det(R_{L-k-1},b_{L-k+1};x,a_{L-k}),\quad \text{if }b_{L-k}=b_{L-k+1}  \\
&(R_{L-k},x)\to \det(R_{L-k-1},a_{L-k+1};x,b_{L-k}),\quad \text{if }a_{L-k}=a_{L-k+1}
    \end{aligned}
    \right.,\label{eq-rep-A}\\
    B_{L}^{(k+1)}=\left\{\begin{aligned}
&B_{L}^{(k)}\x{a_{L-k}}{b_{L-k}}\x{b_{L-k}}{a_{L-k+1}},\quad \text{if }b_{L-k}=b_{L-k+1}  \\
&B_{L}^{(k)}\x{a_{L-k}}{b_{L-k}}\x{a_{L-k}}{b_{L-k+1}},\quad \text{if }a_{L-k}=a_{L-k+1}
    \end{aligned}
    \right|_{\text{eq.(\ref{eq:subs-half})}}\ .\label{eq-rep-B}
\end{gather}
The recursion in $A_L$ can be realized at the level of the $\mathcal A_L$ matrix. Specifically, if $b_{L-k}=b_{L-k+1}$,
\begin{enumerate}
    \item Change the row labels of $\mathcal A_L^{(k)}$ from $\{\dots,R_{L-k}\}$ to $\{\dots,b_{L-k},R_{L-k-1}\}$, and add the column label $a_{L-k+1}$;
    \item Fill in the blanks in the new row and the new column as if the matrix were a Gram matrix; and
    \item Manually set to zero all but the last two rows of the new column, resulting in $\mathcal A_L^{(k+1)}$.
\end{enumerate}
The rules are similar if $a_{L-k}=a_{L-k+1}$. The iteration rules eqs.(\ref{eq:subs-half})-(\ref{eq-rep-B}) apply to any half traintrack integral, not just alternating ones.

Finally, we obtain the 0-form LS of half traintracks:
\begin{equation}
	{\bf LS}^{(L)}_{\rm half}=\frac{{\cal N}_L}{\sqrt{\Delta_{L}}}=\frac{{\cal N}_L}{\sqrt{A_{L}^{2}-4 B_{L}}}\,,
\end{equation}
where $\Delta_L\equiv\Delta_L^{(L)}$, $B_L\equiv B_L^{(L)}$, and $A_{L}\equiv A_L^{(L)}=\det\mathcal{A}_{L}^{(L)}$, which has $2^{L}$ terms. 

\begin{figure}[H]
\centering
	\includegraphics[height=0.2\textwidth]{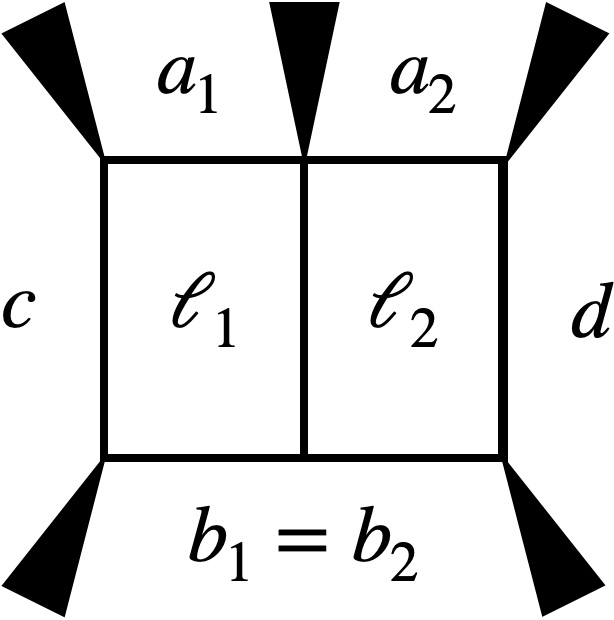}\qquad\includegraphics[height=0.2\textwidth]{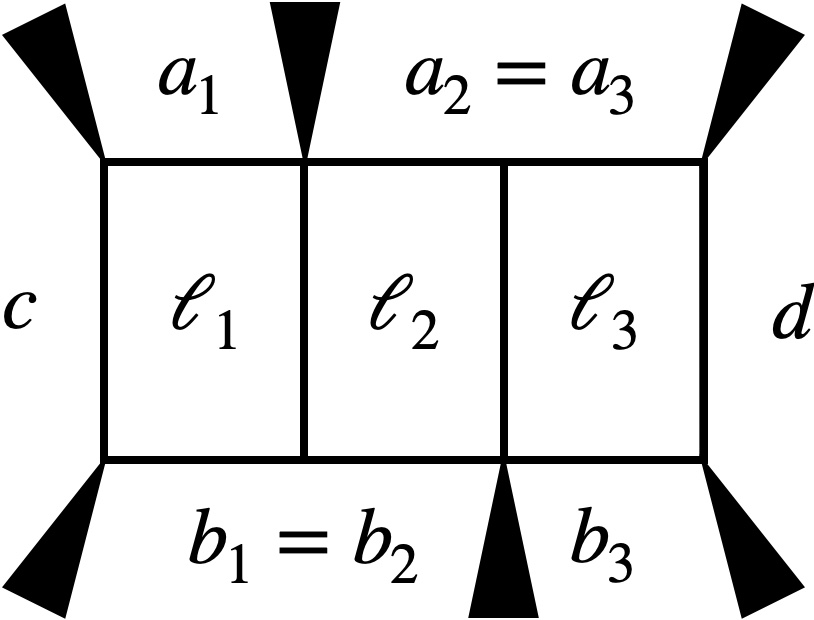}\qquad\includegraphics[height=0.2\textwidth]{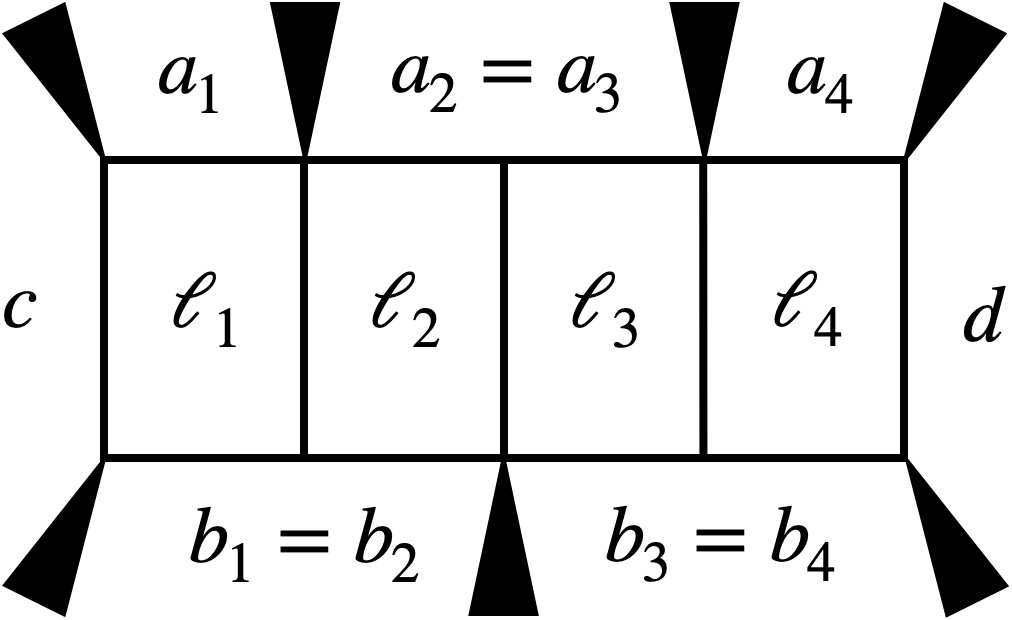}
 \caption{Alternating half traintrack integrals for $L=2,3,4$}
	\label{fig:traintrack_alternating_4loop}
\end{figure}

\paragraph{$L=2,3,4$ alternating traintracks}Consider $\mathfrak T_{\text{alt}}^{(2)}=\left[c\ud{a_1}{b_1}\ud{}{=}\ud{a_2}{b_2}d\right]$ as shown in Fig.~\ref{fig:traintrack_alternating_4loop}. We find that $\Delta_{2}=A_{2}^{2}-4 B_{2}$ where
\begin{equation*}
\begin{aligned}
A_{2}&=\det(\mathcal{A}_{2})=\det\begin{pmatrix}
 		\x{d}{b_{1}}&\x{a_{2}}{b_{1}}&0\\ \x{d}{a_{1}}&\x{a_{2}}{a_{1}}&\x{b_{1}}{a_{1}}\\ \x{d}{c}&\x{a_{2}}{c}&\x{b_{1}}{c}
 		\end{pmatrix}
   \\
   &=\x{b_{1}}{d}\left[\x{a_{1}}{a_{2}}\x{b_{1}}{c}-\x{a_{1}}{b_{1}}\x{a_{2}}{c}\right]-\x{a_{2}}{b_{1}}\left[\x{a_{1}}{d}\x{b_{1}}{c}-\x{a_{1}}{b_{1}}\x{c}{d}\right],\\
   B_{2}&=\x{b_{1}}{c}\x{b_{1}}{d} \x{b_{1}}{a_{1}}\x{b_{1}}{a_{2}} \x{c}{a_{1}}\x{a_{2}}{d}\,.
\end{aligned}
\end{equation*}
Note that $A_{2}$ has $2^{2}=4$ terms and contains $\mathcal N_2=\x{a_{1}}{b_{1}}\x{a_{2}}{b_{1}}\x{c}{d}$. Thus ${\bf LS}^{(2)}_{\rm alt}=\frac{{\cal N}_2}{\sqrt{\Delta_2}}$ is manifestly DCI.

At three loops, $\mathfrak T_{\text{alt}}^{(3)}=\left[c\ud{a_1}{b_1}\ud{}{=}\ud{a_2}{b_2}\ud{=}{}\ud{a_3}{b_3}d\right]$ (Fig.~\ref{fig:traintrack_alternating_4loop}). We have $\Delta_{3}=A_{3}^{2}-4 B_{3}$:
\begin{equation*}
	\begin{aligned}
		A_{3}&=\det\mathcal{A}_{3}=\det\begin{pmatrix}
			\x{d}{a_{2}}&\x{b_{3}}{a_{2}}&0&0\\ \x{d}{b_{1}}&\x{b_{3}}{b_{1}}&\x{a_{2}}{b_{1}}&0\\\x{d}{a_{1}}&\x{b_{3}}{a_{1}}&\x{a_{1}}{a_{2}}&\x{a_{1}}{b_{1}}\\ \x{d}{c}&\x{b_{3}}{c}&\x{a_{2}}{c}&\x{b_{1}}{c} 
		\end{pmatrix}\supset\mathcal N_3,\\
        B_{3}&=\x{b_{1}}{c}\x{b_{3}}{d} \x{b_{1}}{a_{1}}\x{b_{1}}{a_{2}}^{2} \x{c}{a_{1}}\x{a_{2}}{d}\x{a_{2}}{b_{3}}\,.
	\end{aligned}
\end{equation*}

At four loops, $\mathfrak T_{\text{alt}}^{(4)}=\left[c\ud{a_1}{b_1}\ud{}{=}\ud{a_2}{b_2}\ud{=}{}\ud{a_3}{b_3}\ud{}{=}\ud{a_4}{b_4}d\right]$ (Fig.~\ref{fig:traintrack_alternating_4loop}). We have $\Delta_{4}=A_{4}^{2}-4 B_{4}$:
\begin{equation*}
	\begin{aligned}
		A_{4}&=\det\mathcal{A}_{4}=\det\begin{pmatrix}
			\x{d}{b_{3}}&\x{a_{4}}{b_{3}}&0&0&0\\ \x{d}{a_{2}}&\x{a_{4}}{a_{2}}&\x{b_{3}}{a_{2}}&0&0\\\x{d}{b_{1}}&\x{a_{4}}{b_{1}}&\x{b_{1}}{b_{3}}&\x{b_{1}}{a_{2}}&0\\\x{d}{a_{1}}&\x{a_{1}}{a_{4}}&\x{a_{1}}{b_{3}}&\x{a_{1}}{a_{2}}&\x{a_{1}}{b_{1}}\\ \x{d}{c}&\x{a_{4}}{c}&\x{b_{3}}{c}&\x{a_{2}}{c}&\x{b_{1}}{c} 
		\end{pmatrix}\supset\mathcal N_4,\\
        B_4&=\x{b_{1}}{c}\x{b_{3}}{d} \x{b_{1}}{a_{1}}\x{b_{1}}{a_{2}}^{2} \x{c}{a_{1}}\x{a_{4}}{d}\x{a_{2}}{b_{3}}^{2}\x{b_{3}}{a_{4}}\,.
	\end{aligned}
\end{equation*}\hfill$\blacksquare$ 

As already mentioned, the reason for focusing on alternating cases is because the LS of any half traintrack integral is the same as that of a (lower-loop) alternating one. It is straightforward to see this using our method, and we illustrate the reduction with a simple example. 

\begin{figure}[H]
		\centering
	\includegraphics[width=0.4\textwidth]{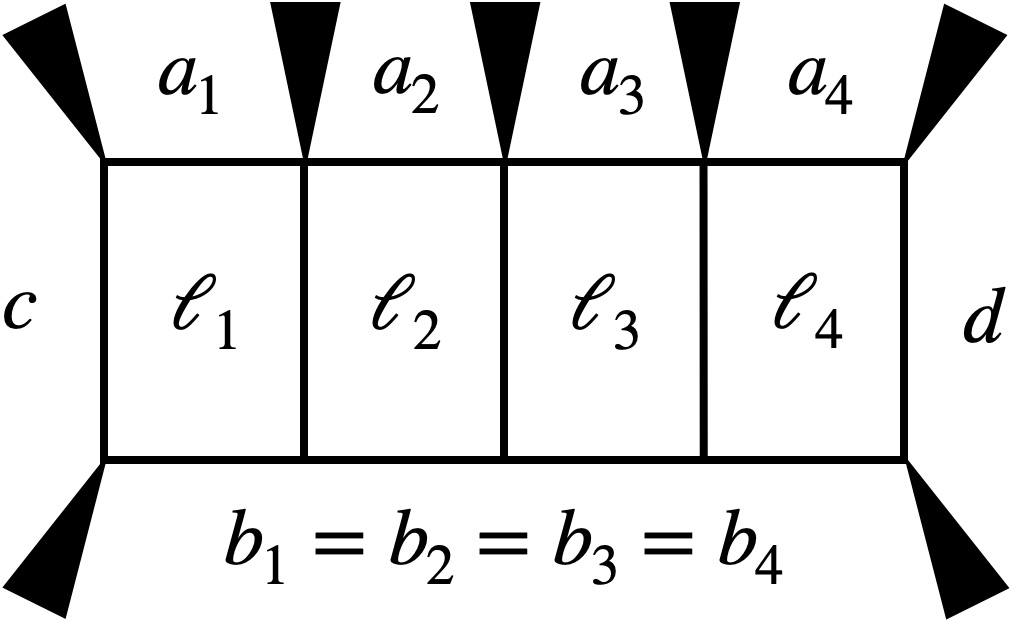}
 \caption{$L=4$ all-up half traintrack integrals}
	\label{fig:traintrack_reduce1}
\end{figure}

\paragraph{$L=4$ all-up traintrack}\label{ex-red-1}Consider $\mathfrak T_{\text{all-up}}^{(4)}=\left[c\ud{a_1}{b_1}\ud{a_2}{b_1}\ud{a_3}{b_1}\ud{a_4}{b_1}d\right]$ as shown in Fig.~\ref{fig:traintrack_reduce1}. We show that the LS of such an all-up traintrack reduce to that of a two-loop alternating one.

On the support of $(R_2,R_2)=0$, we compute the residues at $(R_3,R_3)=(R_4,R_4)=(R_4,d)=0$ corresponding to the two right-most loops to obtain $1/\sqrt{\Delta^{(2)}}$, which equals $\mathbf{LS}\left[R_2\ud{a_3}{b_1}\ud{a_4}{b_1}d\right]$ of a two-looping alternating traintrack. The corresponding $\mathcal A$-matrix has row labels $\{b_1,a_3,R_2\}$ and column labels $\{d,a_4,b_1\}$. Using the recursion rules,
\begin{align*}
    \det\mathcal A^{(3)}&=\x{b_{1}}{a_{3}}\det\begin{pmatrix}
 		\x{d}{b_{1}}&\x{a_{4}}{b_{1}}&0\\ \x{d}{a_{2}}&\x{a_{2}}{a_{4}}&\x{b_{1}}{a_{2}}\\ \x{d}{R_{1}}&\x{a_{4}}{R_{1}}&\x{b_{1}}{R_{1}}
 		\end{pmatrix},\\
   \det\mathcal{A}^{(4)}&=\x{b_{1}}{a_{3}}\x{b_{1}}{a_{2}}\det\begin{pmatrix}
 		\x{d}{b_{1}}&\x{a_{4}}{b_{1}}&0\\ \x{d}{a_{1}}&\x{a_{1}}{a_{4}}&\x{b_{1}}{a_{1}}\\ \x{d}{c}&\x{a_{4}}{c}&\x{b_{1}}{c}
 		\end{pmatrix}.
\end{align*}
Compute $B^{(4)}$ and one finds
\begin{equation*}
    \mathbf{LS}_\text{all-up}\left[c\ud{a_1}{b_1}\ud{a_2}{b_1}\ud{a_3}{b_1}\ud{a_4}{b_1}d\right]=\frac{\mathcal N_4}{\sqrt{\Delta^{(4)}}}=\mathbf{LS}_\text{alt}\left[c\ud{a_1}{b_1}\ud{a_4}{b_1}d\right].
\end{equation*}
The factor $(a_3,b_1)(a_2,b_1)$ is pulled out of $\sqrt{\Delta^{(4)}}$ and cancels part of the numerator $\mathcal N_4$.\hfill$\blacksquare$

\begin{figure}[H]
		\centering
	\includegraphics[width=0.8\textwidth]{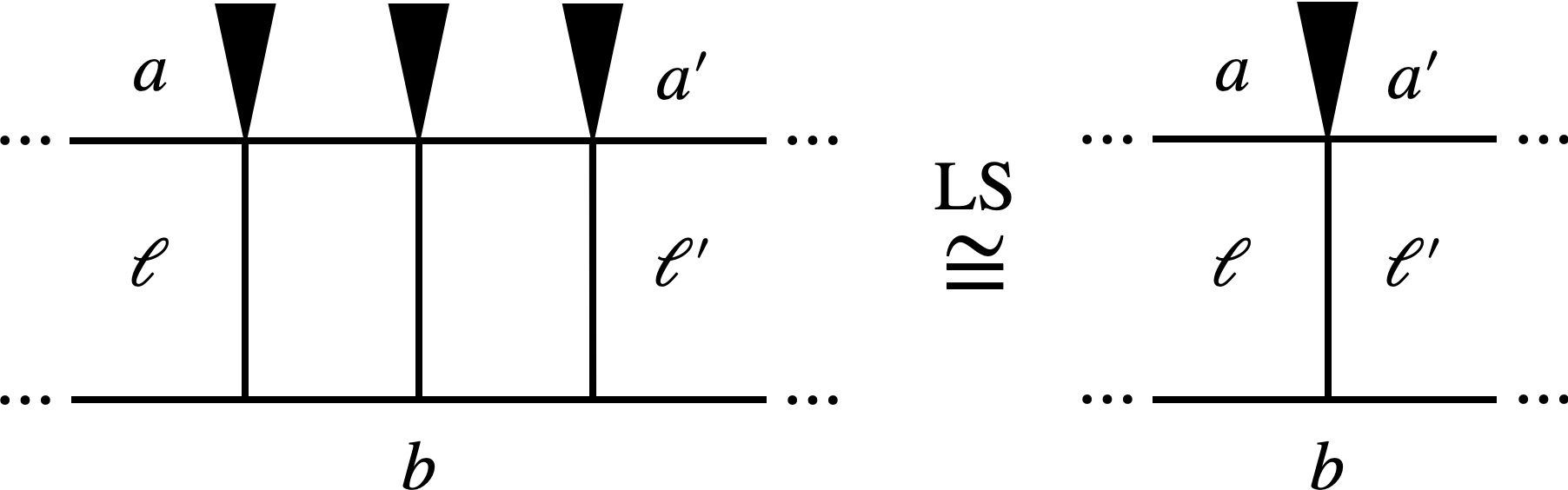}
 \caption{The reduction of LS for half traintrack integrals}
	\label{fig:traintrack_reduce_massive}
\end{figure}

In fact, this generalizes to all half traintracks: the LS can be reduced  
as shown in Fig.~\ref{fig:traintrack_reduce_massive} 
. We will see in the next subsection that such reductions arise in an even simpler way in twistor space.

\subsection{Leading singularities in momentum twistor space}

Momentum twistors~\cite{Hodges:2009hk} are a set of variables that manifest dual conformal invariance and have particularly nice geometric properties; see~\cite{Arkani-Hamed:2010pyv} for a pedagogical introduction. For our purposes, we need only recall that a point $x\in\mathbb{C}^4$ in (complexified) dual space-time corresponds to a line/bi-twistor $(x^1x^2)\subseteq\mathbb{CP}^3$. The choice of $x^1$ and $x^2$ on the line $(x^1x^2)$ is arbitrary, which introduces a $GL(2)$ redundancy. In other words, the space of lines in $\mathbb{CP}^3$ is the Grassmannian $G(2,4)$.

The space-time separation between dual points $x$ and $y$ translate into momentum twistor language as\footnote{We slightly abuse the notation so that $x$ could either mean a dual point (LHS of eq.(\ref{eq:twistor})) or the line/bi-twistor $(x^1x^2)$ (RHS of eq.(\ref{eq:twistor})).}
\begin{equation}
    (x-y)^2=\frac{\langle x^1x^2y^1y^2\rangle}{\langle x^1x^2I_\infty\rangle\langle y^1y^2I_\infty\rangle}\equiv\frac{\langle xy\rangle}{\langle xI_\infty\rangle\langle yI_\infty\rangle}.\label{eq:twistor}
\end{equation}
Here, $I_\infty$ is a bi-twistor that corresponds to the ``space-like infinity'' of the dual space-time, whose presence breaks dual conformal invariance. It must cancel out in any DCI expression. The region variable $\ell_k$ dual to the $k$-th loop momentum corresponds to a line $(A_kB_k)$ in momentum twistor space:
\begin{equation}
    {\rm d}^4\ell_k=\frac{{\rm d}^4A_k{\rm d}^4B_k}{GL(2)}\frac1{\langle A_kB_kI_\infty\rangle^4}.
\end{equation}

In terms of momentum twistors, the traintrack integral reads
\begin{equation}
    \mathfrak T^{(L)}=\left[c\ud{a_1}{b_1}\ud{\cdots}{\cdots}\ud{a_L}{b_L}d\right]=\int\frac{\langle cd\rangle\left(\prod_{k=1}^L\frac{{\rm d}^4A_k{\rm d}^4B_k}{GL(2)}\frac{\langle a_kb_k\rangle}{\langle A_kB_ka_k\rangle\langle A_kB_kb_k\rangle}\right)}{\langle cA_1B_1\rangle\langle A_1B_1A_2B_2\rangle\cdots\langle A_{L-1}B_{L-1}A_LB_L\rangle\langle A_LB_Ld\rangle}.
\end{equation}
Fix the $GL(2)$ redundancies by choosing the parametrization
\begin{equation}
    \left(\begin{matrix}
        A_k\\B_k
    \end{matrix}\right)=\left(\begin{matrix}
        1&0&\alpha_k&\gamma_k\\0&1&\delta_k&\beta_k
    \end{matrix}\right)\left(\begin{matrix}
        a_k^1\\b_k^1\\a_k^2\\b_k^2
    \end{matrix}\right).
\end{equation}
The gauge-fixing introduces a Jacobian~\cite{Arkani-Hamed:2010pyv}:
\begin{equation}
    \left|\frac{\partial(A_kB_k)}{\partial(\alpha_k\beta_k\gamma_k\delta_k)}\right|=\langle a_kb_k\rangle^2.
\end{equation}
We may now trivially take $2L$ residues:
\begin{equation}
    \frac{{\rm d}^4A_k{\rm d}^4B_k}{GL(2)}\frac{\langle a_kb_k\rangle}{\langle A_kB_ka_k\rangle\langle A_kB_kb_k\rangle}=\frac{\langle a_kb_k\rangle^3{\rm d}\alpha_k{\rm d}\beta_k{\rm d}\gamma_k{\rm d}\delta_k}{-\gamma_k\delta_k\langle a_kb_k\rangle^2}\xrightarrow{\mathop{\rm Res}_{\gamma_k,\delta_k=0}}\langle a_kb_k\rangle{\rm d}\alpha_k{\rm d}\beta_k.
\end{equation}
Just to be explicit, on the support of $\gamma_k=\delta_k=0$, the momentum twistors $A_k$ and $B_k$ lie on the lines $a_k=(a_k^1a_k^2)$ and $b_k=(b_k^1b_k^2)$ respectively:
\begin{equation}
    A_k\xlongequal{\gamma_k=\delta_k=0}a_k^1+\alpha_ka_k^2,\quad B_k\xlongequal{\gamma_k=\delta_k=0}b_k^1+\beta_kb_k^2.
\end{equation}

\begin{figure}[H]
    \centering
    \includegraphics[width=0.9\textwidth]{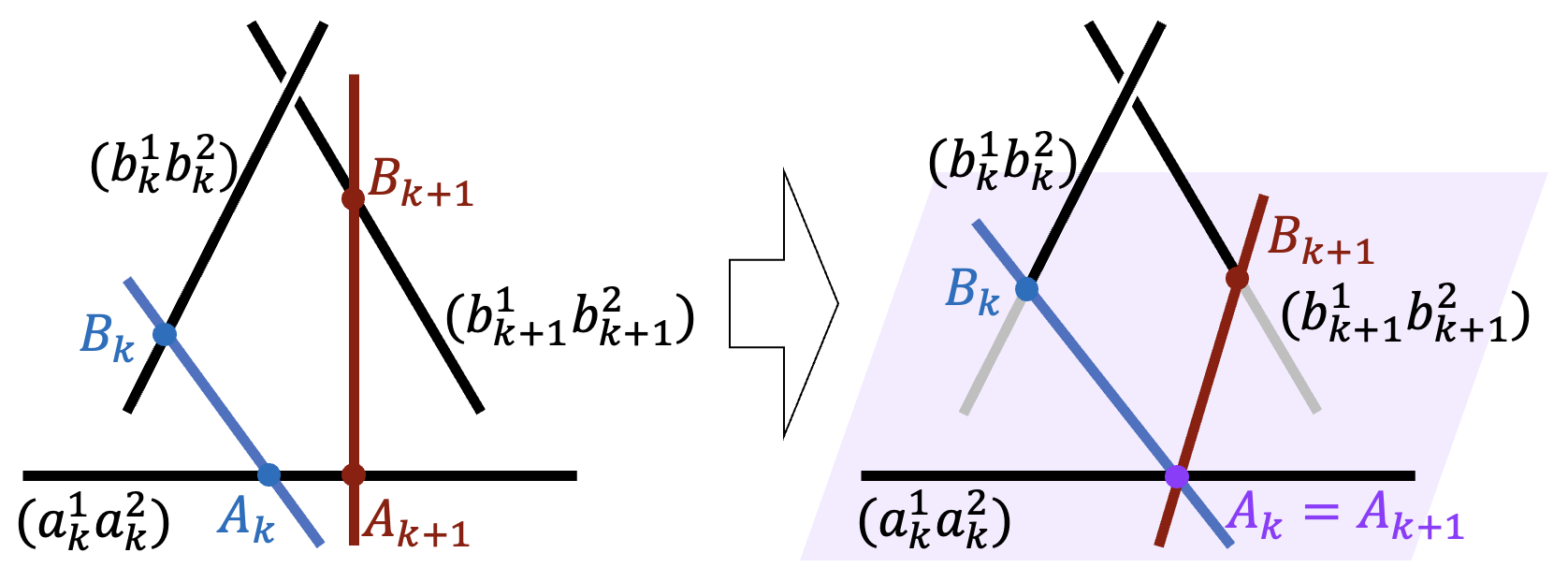}
    \caption{Geometric configuration on the support $\alpha_k-\alpha_{k+1}=\langle B_kB_{k+1} a_k\rangle=0$ of the composite residue}
    \label{fig:composite}
\end{figure}

To take further residues, let us restrict to half traintracks, for which either $a_k=a_{k+1}$ or $b_k=b_{k+1}$ at each loop. An important feature is that $\langle A_kB_kA_{k+1}B_{k+1}\rangle$ factorizes on the support of $\gamma_k=\delta_k=0$, revealing the composite residues. For example, if $a_k=a_{k+1}$,
\begin{equation}
    \langle A_kB_kA_{k+1}B_{k+1}\rangle\xlongequal{\gamma_k=\delta_k=\gamma_{k+1}=\delta_{k+1}=0}(\alpha_k-\alpha_{k+1})\langle B_kB_{k+1}a_k\rangle.
\end{equation}
Therefore, after taking the $2L$ residues with respect to $\{\gamma_k,\delta_k\}_{k=1}^L$, the remaining integrand has exactly $2L$ factors in the denominator, which are
\begin{itemize}
    \item $\langle A_1B_1c\rangle$ and $\langle A_LB_Ld\rangle$ on the left and right ends; and
    \item for each $k=1,\cdots,L{-}1$, either $(\alpha_k-\alpha_{k+1})\langle B_kB_{k+1}a_k\rangle$ or $(\beta_k-\beta_{k+1})\langle A_kA_{k+1}b_k\rangle$.
\end{itemize}
The 0-form LS is supported on the vanishing locus of all these $2L$ factors. Geometrically, $\alpha_k-\alpha_{k+1}=0$ means $A_k=A_{k+1}$, while the non-trivial constraint $\langle B_kB_{k+1}a_k\rangle=0$ means the two planes $(a_kB_k)$ and $(a_kB_{k+1})$ coincide (Fig.~\ref{fig:composite}).

To actually determine the LS, notice that all non-trivial remaining constraints are of the form $\langle EFr\rangle=0$, where $E,F$ are two twistors and $r$ a bi-twistor. Furthermore, $E=e^1+\epsilon e^2$ and $F=f^1+\zeta f^2$ are constrained to live on separate lines $(e^1e^2)$ and $(f^1f^2)$. The parameters $\epsilon$ and $\zeta$ are constrained by a Möbius-type relation:
\begin{equation}
    \langle e^1f^1r\rangle+\epsilon\langle e^2f^1r\rangle+\zeta\langle e^1f^2r\rangle+\epsilon\zeta\langle e^2f^2r\rangle=0\implies\zeta=\frac{-\langle e^2f^1r\rangle\epsilon-\langle e^1f^1r\rangle}{\langle e^2f^2r\rangle\epsilon+\langle e^1f^2r\rangle}.
\end{equation}
It helps to introduce a right-action by a $GL(2)$ matrix
\begin{equation}
    \epsilon\triangleleft M:=\frac{M_{11}\epsilon+M_{21}}{M_{12}\epsilon+M_{22}}.
\end{equation}
The constraint $\langle EFr\rangle=0$ can now be viewed as an unnormalized Möbius transformation ``induced by $r$'':
\begin{equation}
    \zeta=\epsilon\triangleleft M_{\langle EFr\rangle},\quad M_{\langle EFr\rangle}:=\left(\begin{matrix}
        -\langle e^2f^1r\rangle&\langle e^2f^2r\rangle\\-\langle e^1f^1r\rangle&\langle e^1f^2r\rangle
    \end{matrix}\right),\quad\det M_{\langle EFr\rangle}=\langle er\rangle\langle fr\rangle.
\end{equation}
The fact that we define a right-action by $\epsilon\mapsto\frac{M_{11}\epsilon+M_{21}}{M_{12}\epsilon+M_{22}}$ instead of the usual left-action $\epsilon\mapsto\frac{M_{11}\epsilon+M_{12}}{M_{21}\epsilon+M_{22}}$ seems confusing, but we choose this convention so that the transformations compose in the following nice way: (suppose $G=g^1+\eta g^2$)
\begin{equation}
    \langle EFr\rangle=\langle FGr'\rangle=0\implies\eta=\epsilon\triangleleft(M_{\langle EFr\rangle}M_{\langle FGr'\rangle})=:\epsilon\triangleleft M_{\langle EFr\rangle\langle FGr'\rangle}.
\end{equation}

The geometric viewpoint that constraints induce Möbius transformations makes manifest the invariance under reparametrizations. For instance, if we were to redefine $f^1\mapsto f^1+\chi f^2$ (after all, kinematically all that matters is the line $(f^1f^2)$ itself), the meaning of $\zeta$ would change, but the matrix $M_{\langle EFr\rangle\langle FGr'\rangle}$ is invariant:
\begin{equation*}
    M_{\langle EFr\rangle}M_{\langle FGr'\rangle}\mapsto M_{\langle EFr\rangle}\left(\begin{matrix}
        1\\-\chi&1
    \end{matrix}\right)\left(\begin{matrix}
        1\\\chi&1
    \end{matrix}\right)M_{\langle FGr'\rangle}=M_{\langle EFr\rangle}M_{\langle FGr'\rangle}.
\end{equation*}
The deeper reason is that such a reparametrization merely redefines the coordinate $\zeta$ on the line/Riemann sphere $(f^1f^2)\cong\mathbb{CP}^1$, which corresponds to a Möbius transformation $\zeta\mapsto\zeta\triangleleft X$. Then, $M_{\langle EFr\rangle}\mapsto M_{\langle EFr\rangle}X$ and $M_{\langle FGr'\rangle}\mapsto X^{-1}M_{\langle FGr'\rangle}$. The same is true for other $\mathbb{CP}^1$-isomorphisms as well, such as $f^1+\zeta f^2\mapsto(1-\zeta)f^1+\zeta f^2$.

For half traintracks, the $(L+1)$ constraints of the form $\langle EFr\rangle=0$ constitute a cycle with ``monodromy''
\begin{equation}
    \mathcal M:=M_{\langle EFr\rangle\cdots\langle GEr'\rangle}.\label{eq:cycle}
\end{equation}
In the end, we would obtain a self-consistency equation
\begin{equation}
    \epsilon=\frac{\mathcal M_{11}\epsilon+\mathcal M_{21}}{\mathcal M_{12}\epsilon+\mathcal M_{22}}\implies\mathcal M_{12}\epsilon^2+(\mathcal M_{22}-\mathcal M_{11})\epsilon-\mathcal M_{21}=0,
\end{equation}
whose discriminant reads
\begin{equation}
    \Delta_L=(\mathcal M_{22}-\mathcal M_{11})^2+4\mathcal M_{12}\mathcal M_{21}=(\mathop{\rm tr}\mathcal M)^2-4\det\mathcal M.
\end{equation}
By following the residue-taking process carefully and keeping track of the prefactors, it is straightforward to show that the leading singularity of the traintrack integral is
\begin{equation}
    \mathbf{LS}^{(L)}_\text{half}=\frac{\langle cd\rangle\langle a_1b_1\rangle\cdots\langle a_Lb_L\rangle}{\sqrt{\Delta_L}}.\label{eq:twistorLS}
\end{equation}

Incidentally, since $\mathcal M$ is the product of a string of $2\times2$ matrices, both $\mathop{\rm tr}\mathcal M$ and $\det\mathcal M$ are invariant under a cyclic rotation of the individual $M_{\langle EFr\rangle}$ matrices, which makes it clear that $\sqrt{\Delta_L}$ is the same no matter which line parameter $\epsilon$ we choose to represent it.

\paragraph{$L=3$ alternating traintrack}Consider $\mathfrak T_\text{alt}^{(3)}=\left[c\ud{a}{b}\ud{a'}{b}\ud{a'}{b'}d\right]$, with constraints $\langle B_1A_1c\rangle=\langle A_1A_2b\rangle=\langle A_2{\color{blue}(=A_3)}B_3d\rangle=\langle B_3B_1{\color{blue}(=B_2)}a'\rangle=0$. Hence,
\begin{equation*}
    \det\mathcal M=\langle bc\rangle\langle ac\rangle\times\langle ab\rangle\langle a'b\rangle\times\langle a'd\rangle\langle b'd\rangle\times\langle b'a'\rangle\langle ba'\rangle.
\end{equation*}
Simple exercises like this provide a consistency check that eq.(\ref{eq:twistorLS}) is indeed DCI.\hfill$\blacksquare$

\paragraph{$L=3$ all-up traintrack}\label{ex-red-2}Consider $\mathfrak T^{(3)}_\text{all-up}=\left[c\ud{a_1}{b}\ud{a_2}{b}\ud{a_3}{b}d\right]$, with constraints $\langle B_1A_1c\rangle=\langle A_1A_2b\rangle=\langle A_2A_3b\rangle=\langle A_3B_1{\color{blue}(=B_3)}d\rangle=0$. Now, using Schouten identities, it is straightforward to show that
\begin{equation}
    M_{\langle A_1A_2b\rangle\langle A_2A_3b\rangle}=M_{\langle A_1A_3b\rangle}\langle a_2b\rangle.\label{eq:shorten}
\end{equation}
Therefore, $\mathcal M_\text{original}\propto\mathcal M_\text{reduced}$, and the proportionality factor $\Delta_\text{original}/\Delta_\text{reduced}=\langle a_2b\rangle^2$ factors out of $\sqrt{\Delta_L}$ to cancel the $\langle a_2b\rangle$ in the numerator of eq.(\ref{eq:twistorLS}). As far as the LS is concerned, the all-up traintrack is equivalent to the reduced alternating version:
\begin{equation}
    \mathbf{LS}_\text{all-up}\left[c\ud{a_1}{b}\ud{a_2}{b}\ud{a_3}{b}d\right]=\mathbf{LS}_\text{alt}\left[c\ud{a_1}{b}\ud{a_3}{b}d\right].
\end{equation}
Geometrically, this is understood as coincidence/coplanarity being transitive.\hfill$\blacksquare$

The same reasoning shows that any traintrack integral\footnote{The reduction eq.(\ref{eq:shorten}) applies to any part of a traintrack integral, even those with higher-form LS.} has the same LS as its reduced version, with each $a_k$ or $b_k$ opposite to at most two distinct dual points (Fig.~\ref{fig:traintrack_reduce_massive}). This generalizes the well-known fact that a ladder $\left[c\ud{a}{b}\ud{a}{b}\ud{\cdots}{\cdots}\ud{a}{b}d\right]$ has the same LS as the box itself.

There is a recursive structure to the ``monodromy'' $\mathcal M$. To see this, consider two traintracks $\left[c\ud{\cdots}{\cdots}\ud{a_{L-1}}{b_{L-1}}d\right]$ and $\left[c\ud{\cdots}{\cdots}\ud{a_{L-1}}{b_{L-1}}\ud{}{=}\ud{a_L}{b_L}d\right]$. Accordingly,
\begin{equation}
    \mathcal M_{L-1}=M_{\langle A_{L-1}B_{L-1}d\rangle\cdots}=\left(\begin{matrix}
        -\langle a_{L-1}^2b_L^1d\rangle&\langle a_{L-1}^2b_L^2d\rangle\\-\langle a_{L-1}^1b_L^1d\rangle&\langle a_{L-1}^1b_L^2d\rangle
    \end{matrix}\right)M_{\cdots}
\end{equation}
and
\begin{align*}
    \mathcal M_L&=M_{\langle A_{L-1}A_Lb_L\rangle\langle A_LB_{L-1}d\rangle\cdots}\\
    &=\left(\begin{matrix}
        -\langle a_{L-1}^2a_L^1b_L\rangle&\langle a_{L-1}^2a_L^2b_L\rangle\\-\langle a_{L-1}^1a_L^1b_L\rangle&\langle a_{L-1}^1a_L^2b_L\rangle
    \end{matrix}\right)\left(\begin{matrix}
        -\langle a_L^2b_L^1d\rangle&\langle a_L^2b_L^2d\rangle\\-\langle a_L^1b_L^1d\rangle&\langle a_L^1b_L^2d\rangle
    \end{matrix}\right)M_{\cdots}\\
    &=\left(\begin{matrix}
        \langle a_{L-1}^2a_L^1b_L\rangle\langle {\color{blue}a_L^2}b_L^1d\rangle-\langle a_{L-1}^2a_L^2b_L\rangle\langle {\color{blue}a_L^1}b_L^1d\rangle&\ast\\\ast&\ast
    \end{matrix}\right)M_{\cdots}.
\end{align*}
To save space, we have omitted the other three entries which have a similar expression. 

Using the Schouten identity on the blue twistor,
\begin{equation*}
    \langle a_{L-1}^2a_L^1b_L\rangle\langle {\color{blue}a_L^2}b_L^1d\rangle-\langle a_{L-1}^2a_L^2b_L\rangle\langle {\color{blue}a_L^1}b_L^1d\rangle=\langle a_{L-1}^2b_L^1a_L\rangle\langle b_Ld\rangle-\langle a_Lb_L\rangle\langle a_{L-1}^2b_L^1d\rangle.
\end{equation*}
Repeat this for the other three entries. In the end, we obtain
\begin{align}
    \mathcal M_L&=[-\langle b_Ld\rangle M_{\langle A_{L-1}B_{L-1}a_L\rangle}+\langle a_Lb_L\rangle M_{\langle A_{L-1}B_{L-1}d\rangle}]M_{\cdots}\nonumber\\
    &=\langle a_Lb_L\rangle\mathcal M_{L-1}-\langle b_Ld\rangle\mathcal M_{L-1}|_{d\to a_L}.
\end{align}

The recursion shows that $\mathop{\rm tr}\mathcal M_L$ depends on the dual points only, because $\mathop{\rm tr}\mathcal M_1=\langle cd\rangle\langle a_1b_1\rangle-\langle a_1c\rangle\langle b_1d\rangle-\langle b_1c\rangle\langle a_1d\rangle$ depends on dual points only for the box integral $\left[c\ud{a_1}{b_1}d\right]$. In fact, this is exactly the same recursion rule as that of $A_L$ in the Feynman parametrization approach, as can be seen by Laplace-expanding the determinant in eq.(\ref{eq:amat2}) with respect to the first line.

\section{Higher-form leading singularities of general traintrack integrals and Calabi-Yau geometries}\label{sec-ls-cy}

\begin{figure}[H]
    \centering
    \includegraphics[width=0.3\textwidth]{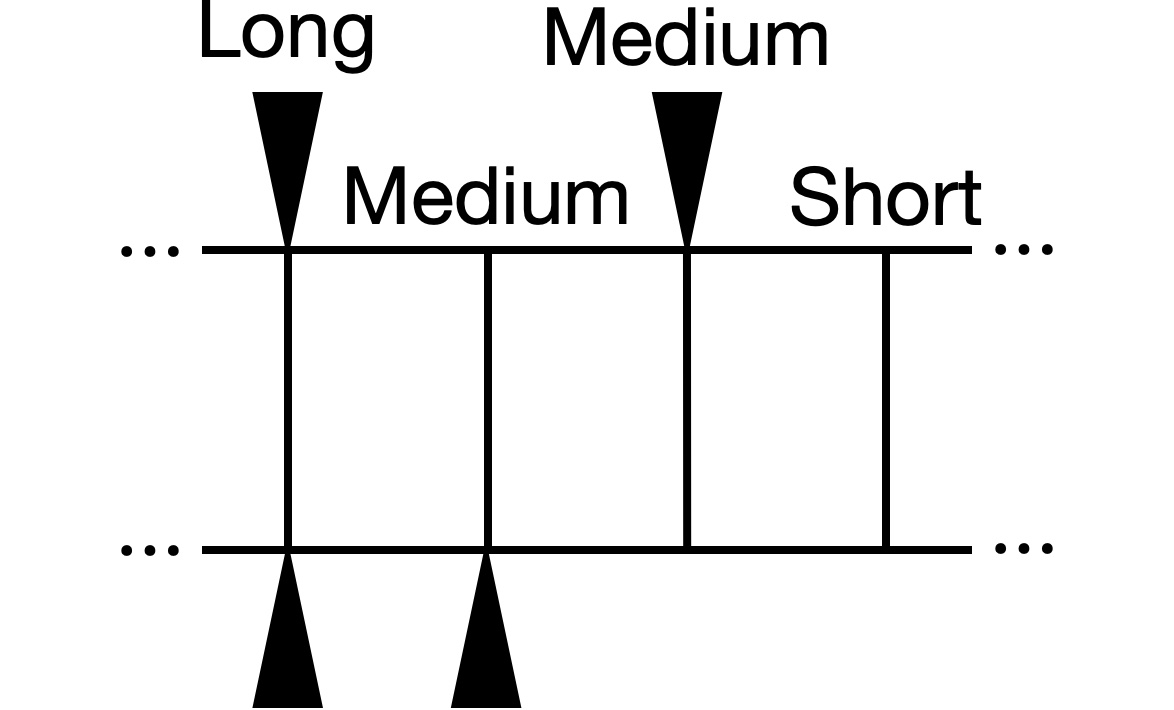}
    \caption{Three types of rungs along the traintrack}
    \label{fig:ties}
\end{figure}

In this section, we move to the computation of higher-form LS for general (massive) traintrack integrals. Whenever there exists ``long'' rungs (Fig.~\ref{fig:ties}), \emph{i.e.}, neighboring pairs of dual points such that $a_k\neq a_{k+1}$ and $b_k\neq b_{k+1}$, some composite residues disappear. As a consequence, the maximal codimension residue (the generalized notion of LS as proposed in~\cite{Bourjaily:2020hjv}) has codimension $<2L$.

In particular, we will study the full traintrack with no medium or short rungs (all $\{a_k,b_k\}_{k=1}^L$ are distinct). Here, one can take $L{+}1$ residues in Feynman parameter space, or $3L{+}1$ residues in momentum (twistor) space, resulting in the integral 
of an $(L{-}1)$-form. Surprisingly, these two seemingly unrelated parametrization lead to similar results. For instance, both yield an $(L{-}1)$-fold integral of the square root of an irreducible polynomial, quartic with respect to each of the $L{-}1$ variables. Along the way, we consider the somewhat more invariant notion of ``sub-leading singularities'' (subLS), which is an $L$-form in $(\mathbb{CP}^1)^L$. The $(L{-}1)$-form LS is nothing but the residue of the $L$-form subLS. The subLS has a simple singularity on the vanishing hypersurface quadratic in each of the $L$ variables. It then follows\footnote{For a pedagogical introduction to Calabi-Yau manifolds, see, for example,~\cite{Bouchard:2007ik}.} that the hypersurface is a Calabi-Yau manifold (possibly with singularities), and the LS its unique holomorphic top-form.

Several remarks are now in order. As defined in~\cite{Bourjaily:2020hjv}, the LS is a period integral, which means that in principle we need to specify an integration contour to pair with the rigid $(L{-}1)$-form integrand obtained by taking residues. However, we decide to focus on the integrand only, leaving a study of contours to future work. As a result, our presentation of higher-form LS will have a certain arbitrariness, as we are free to choose whichever variables to parametrize the differential form. Our convention prefers $\alpha$-variables to $\beta$-variables, but the differential form itself is invariant under reparametrization.

We will mostly focus on full traintracks in this section, but will comment briefly on the less rigid cases. Full traintracks have $R=L-1$ long rungs, while half traintracks have $R=0$ long rungs. Less rigid traintracks have $0<R<L-1$ long rungs, \emph{i.e.}, there are $R$ values of $k$ such that $a_k\neq a_{k+1}$ and $b_k\neq b_{k+1}$, which we denote by $k_1,\cdots,k_R$.
In general, they can be viewed either as the degeneration of an $L$-loop full traintrack under soft limits, or as the modification of an $(R+1)$-loop full traintrack with medium and short rungs inserted. The details of how such degenerations and modifications take place are discussed in the following. In short, such a traintrack integral has an $R$-form LS.

\subsection{Higher-form leading singularities in Feynman parameter space} \label{sec-highform-fey}

Recall that the massive full traintrack is defined by eqs.(\ref{inital_feynman_rep})-(\ref{eq-gl1}), with $GL(1)^L$ redundancies. As mentioned above, we expect the LS to be an $(L{-}1)$-form. To compute the codim-$(L{+}1)$ residue, we need convenient choices of gauge-fixing and parametrization. The authors of~\cite{Bourjaily:2018ycu} chose to fix $\gamma_k=1$ and parametrized the $(L{-}1)$-form with $\{\alpha_k\}_{k=1}^{L-1}$, resulting in a differential form with a non-trivial numerator. 
Here, we make a different choice. We leave the $GL(1)^L$ invariance intact, and take residues with respect to the variables $\gamma_k$ on the rungs. We shall see that this leads to a representation of the higher-form LS associated with a CY manifold embedded in $(\mathbb{CP}^{1})^L$. Moreover, such a representation appears symmetric with respect to the $L$ loops.


The basic strategy is similar to the half traintrack case. First, we take residues using factors from the right-most loop, to obtain a basic structure. Then, as we iteratively take further residues, such a structure gets deformed.

We start by taking the residue at $(R_L,R_L)=0$ with respect to $\gamma_L$, on the support of $(R_{L-1},R_{L-1})=0$.
\begin{equation}
    (R_L,R_L)=0\implies\gamma_L=\frac{-\alpha_L\beta_L(a_L,b_L)}{\alpha_L(a_L,R_{L-1})+\beta_L(b_L,R_{L-1})}\equiv\gamma_L^*.
\end{equation}
A straightforward calculation yields
\begin{equation}
    \mathop{\rm Res}_{(R_L,R_L)=0}\frac{{\rm d}\gamma_L{\rm d}\alpha_L{\rm d}\beta_L}{(R_L,R_L)(R_L,d)}=\left.\frac{1}{(R_L,d)}\right|_{\gamma_L=\gamma_L^*}\frac{{\rm d}\alpha_L{\rm d}\beta_L}{\mathcal J},
\end{equation}
where the Jacobian is $\mathcal J=\alpha_L(a_L,R_{L-1})+\beta_L(b_L,R_{L-1})$, and
\begin{equation}
    \begin{aligned}
        \left.(R_L,d)\right|_{\gamma_L=\gamma_L^*}=\frac{\begin{pmatrix}
            \alpha_L&\beta_L
        \end{pmatrix}\mathbb Q_d^L\begin{pmatrix}
            \alpha_L\\\beta_L
        \end{pmatrix}}{\alpha_L(a_L,R_{L-1})+\beta_L(b_L,R_{L-1})}.
    \end{aligned}
\end{equation}
We will frequently run into quadrics like $\mathbb Q_d^L$ above. The general definition is
\begin{equation}\label{eq-quar}
	\mathbb{Q}_{x}^{k}\equiv\begin{pmatrix}
		\x{x}{a_{k}}\x{a_{k}}{R_{k-1}}&\frac12A(a_{k},R_{k-1},b_{k},x)\\ \frac12A(a_{k},R_{k-1},b_{k},x)&\x{x}{b_{k}}\x{b_{k}}{R_{k-1}}
	\end{pmatrix},
\end{equation}
where we $A(a,b,c,d)\equiv\x{a}{b}\x{c}{d}+\x{a}{d}\x{b}{c}-\x{a}{c}\x{b}{d}$. In the end, the Jacobian cancels the denominator of $(R_L,d)$, yielding the net result:
\begin{equation}
    \mathop{\rm Res}_{(R_L,R_L)=0}\frac{{\rm d}\gamma_L{\rm d}\alpha_L{\rm d}\beta_L}{(R_L,R_L)(R_L,d)}=\frac{{\rm d}\alpha_L{\rm d}\beta_L}{\mathcal D_L^{(1)}},\quad\mathcal D_L^{(1)}\equiv\vec\alpha_L\cdot\mathbb Q_d^L\cdot\vec\alpha_L,\quad\vec\alpha_L\equiv(\alpha_L,\beta_L).
\end{equation}

Taking further residues is completely straightforward, once we realize that $\mathcal D_L^{(1)}$ is linear in $(R_{L-1})$ so that the Jacobian always cancels the denominator. For instance,
\begin{gather}
    \mathop{\rm Res}_{(R_{L-1},R_{L-1})=0}\frac{{\rm d}\gamma_{L-1}{\rm d}\alpha_{L-1}{\rm d}\beta_{L-1}}{(R_{L-1},R_{L-1})\mathcal D_L^{(1)}}=\frac{{\rm d}\alpha_{L-1}{\rm d}\beta_{L-1}}{\mathcal D_L^{(2)}},\\
    \mathcal D_L^{(2)}\equiv\mathcal D_L^{(1)}\Big|_{(R_{L-1},x)\to\vec\alpha_{L-1}\cdot\mathbb Q_x^{L-1}\cdot\vec\alpha_{L-1}}.
\end{gather}
In this way, we could easily take $L$ residues to obtain the sub-leading singularity:
\begin{equation}
    \mathbf{subLS}^{(L)}_\text{full}=\int\left(\prod_{k=1}^L\frac{{\rm d}\alpha_k{\rm d}\beta_k}{GL(1)}\right)\frac{\mathcal N_L}{\mathcal D_L},
\end{equation}
where $\mathcal D_L$ is an irreducible quadratic polynomial with respect to each $\alpha_k$:
\begin{gather}
    (R_k,x)\to\vec\alpha_k\cdot\mathbb Q_x^k\cdot\vec\alpha_k,\label{eq:subs-full}\\
    \mathcal D_L\equiv\mathcal D_L^{(L)}=(R_L,d)\Big|_{\text{eq.(\ref{eq:subs-full})}}.
\end{gather}

By now, it is clear that the $\mathcal D_L$ is quadratic with respect to the parameters from all $L$ loops. Moreover, it has a simple singularity on the CY hypersurface $\mathcal D_L=0$ embedded in $(\mathbb{CP}^1)^L$. The leading singularity is obtained by taking one further residue, which yields the unique holomorphic top-form of the CY manifold.
\begin{equation}
    \mathbf{LS}^{(L)}_\text{full}=\oint\mathbf{subLS}^{(L)}_\text{full}.
\end{equation}
For instance, we could present it by solving for $\alpha_L$ for the final residue, and fixing the projective invariance by setting all $\beta$'s to be unity:
\begin{equation}
    \mathbf{LS}^{(L)}_\text{full}=\mathcal N_L\int\frac{{\rm d}\alpha_1\cdots{\rm d}\alpha_{L-1}}{\sqrt{\mathcal Q_L(\alpha_1,\cdots,\alpha_{L-1})}}.
\end{equation}
Here, $\mathcal Q_L$ is an irreducible quartic with respect to each $\alpha_k$.

The CY manifold can be put into elliptic-fibered form, by transforming $\mathcal Q_L$ into Weierstrass form with respect to any $\alpha_k$. To achieve this, write
\begin{equation}
    \mathcal Q_L(\alpha)=\sum_{i=0}^4c_i(\widehat{\alpha_k})\alpha_k^i\implies y^2=4x^3-g_2(\widehat{\alpha_k})x-g_3(\widehat{\alpha_k}),
\end{equation}
where $\widehat{\alpha_k}$ denotes all $\alpha$'s except $\alpha_k$. The coefficient functions $g_2(\widehat{\alpha_k})$ and $g_3(\widehat{\alpha_k})$ have degrees 8 and 12 with respect to each variable.

For less rigid traintracks, there are two perspectives as mentioned earlier. From the first perspective, whenever a long rung shortens, say the one between the $k_r$- and the $(k_r{+}1)$-th loops, the subLS $\mathcal D_L$ factorizes into two factors, each linear with respect to $\alpha_{k_r}$ and $\alpha_{k_r+1}$, revealing a composite residue. Hence, for every medium or short rung, the rigidity of LS decreases by one, which implies a traintrack with $R$ long rungs has an $R$-form LS.

From the second perspective, we start from the right-most loop as always, but apply different iteration rules according to whether a rung is long or not. Whenever we encounter a long rung, we use the rule in eq.(\ref{eq:subs-full}), whereas for a medium or short rung, we use the rule in eq.(\ref{eq:subs-half}). It is obvious that, between long rungs, the medium and short rungs induce the rule eq.(\ref{eq:subs-half}) which merely deforms the entries of the quadric $\mathbb{Q}_{x}^{k_r}$. Therefore, the rigidity is unchanged by the insertions of medium or short rungs, which proves the fact that a traintrack with $R$ long rungs has an $R$-form LS. We present an example of the $R=1$ case in appendix~\ref{app-detail-ell}.

\subsection{Higher-form leading singularities in momentum twistor space}

To determine the higher-form LS using momentum twistors, let us first analyze the constraints imposed by cutting propagators. As with the 0-form LS case, the conditions $\langle A_kB_ka_k\rangle=\langle A_kB_kb_k\rangle=0$ are trivially solved by the parametrization $A_k=a_k^1+\alpha_ka_k^2$ and $B_k=b_k^1+\beta_kb_k^2$. However, $\langle A_kB_kA_{k+1}B_{k+1}\rangle$ no longer factorizes if $a_k\neq a_{k+1}$ and $b_k\neq b_{k+1}$, which means there are now fewer factors in the denominator, making it impossible to take another $2L$ residues. In the extreme case of full traintracks, there are only $(L{+}1)$ factors, one for each rung: (recall that $\ell_k=(A_kB_k)$)
\begin{equation}
    \langle\ell_0\ell_1\rangle=\langle\ell_1\ell_2\rangle=\cdots=\langle\ell_L\ell_{L+1}\rangle=0,\quad\ell_0\equiv c,\quad\ell_{L+1}\equiv d.\label{eq:constraints}
\end{equation}
Therefore, the residue-taking process terminates early, leaving us with a rigid $(L{-}1)$-form LS.



Let us first focus on full traintracks, whose LS is expected to be a rigid $(L{-}1)$-form. To compute it, take the residue at $\langle\ell_1\ell_2\rangle=0$ with respect to $\beta_1$:
\begin{gather}
    \langle A_1B_1\ell_2\rangle=0\implies\beta_1=\alpha_1\triangleleft M^{(1)},\quad M^{(1)}\equiv M_{\langle A_1B_1\ell_2\rangle},\label{eq:shortcycle1}\\
    \mathop{\rm Res}_{\langle\ell_1\ell_2\rangle=0}\frac{{\rm d}\alpha_1{\rm d}\beta_1}{\langle\ell_0\ell_1\rangle\langle\ell_1\ell_2\rangle}=\left.\frac1{\langle\ell_1c\rangle}\right|_{\beta_1=\alpha_1\triangleleft M^{(1)}}\frac{{\rm d}\alpha_1}{\mathcal J}.
\end{gather}
The Jacobian reads
\begin{equation}
    \mathcal J=\frac{\partial\langle\ell_1\ell_2\rangle}{\partial\beta_1}=\langle A_1b_1^2\ell_2\rangle=M^{(1)}_{22}+M^{(1)}_{12}\alpha_1.
\end{equation}
On the other hand,
\begin{equation*}
    \langle\ell_1c\rangle\xlongequal{\beta_1=\alpha_1\triangleleft M^{(1)}}\frac{\left[M^{(1)}_{22}+M^{(1)}_{12}\alpha_1\right]\langle A_1b_1^1c\rangle+\left[M^{(1)}_{21}+M^{(1)}_{11}\alpha_1\right]\langle A_1b_1^2c\rangle}{M^{(1)}_{22}+M^{(1)}_{12}\alpha_1}.
\end{equation*}
Since $A_1=a_1^1+\alpha_1a_1^2$, the numerator is a quadratic polynomial in $\alpha_1$:
\begin{equation}
    \langle\ell_1c\rangle\xlongequal{\beta_1=\alpha_1\triangleleft M^{(1)}}\frac{D_0+D_1\alpha_1+D_2\alpha_1^2}{M^{(1)}_{22}+M^{(1)}_{12}\alpha_1}=:\frac{\mathcal D_L^{(1)}}{M^{(1)}_{22}+M^{(1)}_{12}\alpha_1}.
\end{equation}
When the smoke clears,
\begin{equation}
    \mathop{\rm Res}_{\langle\ell_1\ell_2\rangle=0}\frac{{\rm d}\alpha_1{\rm d}\beta_1}{\langle\ell_0\ell_1\rangle\langle\ell_1\ell_2\rangle}=\frac{{\rm d}\alpha_1}{\mathcal D_L^{(1)}}.
\end{equation}

Notice that the matrix elements of $M^{(1)}\equiv M_{\langle A_1B_1\ell_2\rangle}$ and hence $D_i$ and $\mathcal D_L^{(1)}$ are proportional to the bi-twistor $(\ell_2)$. This key feature enables us to iteratively take more residues. For example, to take the residue at $\langle\ell_2\ell_3\rangle=0$ with respect to $\beta_2$,
\begin{gather}
    \langle A_2B_2\ell_3\rangle=0\implies\beta_2=\alpha_2\triangleleft M^{(2)},\quad M^{(2)}\equiv M_{\langle A_2B_2\ell_3\rangle},\label{eq:shortcycle2}\\
    \mathop{\rm Res}_{\langle\ell_2\ell_3\rangle=0}\frac{{\rm d}\alpha_2{\rm d}\beta_2}{\mathcal D_L^{(1)}\langle\ell_2\ell_3\rangle}=\left.\frac1{\mathcal D_L^{(1)}}\right|_{\beta_2=\alpha_2\triangleleft M^{(2)}}\frac{{\rm d}\alpha_2}{\mathcal J}.
\end{gather}
Since $\mathcal D_L^{(1)}$ is proportional to $(\ell_2)=(a_2^1b_2^1)+\alpha_2(a_2^2b_2^1)+\beta_2(a_2^1b_2^2)+\alpha_2\beta_2(a_2^2b_2^2)$,
\begin{equation}
    \mathcal D_L^{(1)}\xlongequal{\beta_2=\alpha_2\triangleleft M^{(2)}}\frac{D_0'+D_1'\alpha_2+D_2'\alpha_2^2}{M^{(2)}_{22}+M^{(2)}_{12}\alpha_2}=:\frac{\mathcal D_L^{(2)}}{M^{(2)}_{22}+M^{(2)}_{12}\alpha_2}.
\end{equation}
Once again, the denominator cancels the Jacobian, leaving
\begin{equation}
    \mathop{\rm Res}_{\langle\ell_2\ell_3\rangle=0}\frac{{\rm d}\alpha_2{\rm d}\beta_2}{\mathcal D_L^{(1)}\langle\ell_2\ell_3\rangle}=\frac{{\rm d}\alpha_2}{\mathcal D_L^{(2)}}.
\end{equation}
Here, $\mathcal D_L^{(2)}$ is quadratic with respect to both $\alpha_1$ and $\alpha_2$. Moreover, it is proportional to $(\ell_3)$!

By now, it is clear that we can iteratively take $L$ residues with respect to $\{\beta_k\}_{k=1}^L$. The result is
\begin{equation}
    \mathbf{subLS}^{(L)}_\text{full}=\int{\rm d}\alpha_1\cdots{\rm d}\alpha_L\frac{\langle cd\rangle\langle a_1b_1\rangle\cdots\langle a_Lb_L\rangle}{\mathcal D_L},\quad \mathcal D_L\equiv\mathcal D_L^{(L)}=\sum_{i_1,\cdots,i_L=0}^2D_{i_1\cdots i_L}\alpha_1^{i_1}\cdots\alpha_L^{i_L}.
\end{equation}
Because $\mathcal D_L$ is an irreducible quadratic with respect to each $\alpha_k$, when we take the final residue (say, with respect to $\alpha_L$), we would obtain a square root of an irreducible quartic. Specifically,
\begin{gather}
    \mathbf{LS}^{(L)}_\text{full}=\oint\mathbf{subLS}^{(L)}_\text{full}=\int{\rm d}\alpha_1\cdots{\rm d}\alpha_{L-1}\frac{\langle cd\rangle\langle a_1b_1\rangle\cdots\langle a_Lb_L\rangle}{\sqrt{\mathcal Q_L}},\\
    \mathcal Q_L=\sqrt{\left([\alpha_L^1]\mathcal D_L\right)^2-4\left([\alpha_L^0]\mathcal D_L\right)\left([\alpha_L^2]\mathcal D_L\right)},
\end{gather}
where $[\alpha_L^i]\mathcal D_L$ means the coefficient function of $\alpha_L^i$ in the polynomial $\mathcal D_L$. We emphasize that we could have taken the final residue with respect to other $\alpha_k$, or even represent the LS by $\beta$'s instead of $\alpha$'s; the differential form is independent of these choices.




As for less rigid traintracks, from the first perspective, each time a long rung shortens (i.e., $a_k=a_{k+1}$ or $b_k=b_{k+1}$, or both) under the soft limit, the polynomial $\mathcal D_L$ becomes reducible with respect to $\alpha_k$ and $\alpha_{k+1}$. Specifically, $\mathcal D_L$ factorizes into two factors that are linear with respect to $\alpha_k$ and $\alpha_{k+1}$, which enables us to take a further residue. For example, if $a_k=a_{k+1}$, one of the factors would be $(\alpha_k-\alpha_{k+1})$. Accordingly, the degree of subLS and hence that of LS drop by one. An $L$-loop traintrack with $R$ long rungs has rigidity $R$.

From the second perspective, recall that $(R+1)$-loop full traintrack satisfies $R+2$ constraints (see eq.(\ref{eq:constraints})), which organize into $R$ overlapping cycles of constraints in the sense of eq.(\ref{eq:cycle}):
\begin{equation}
    \begin{aligned}
        \cdots\qquad\quad&\\
        \langle A_kB_k\ell_{k-1}\rangle=&\,\langle B_kA_k\ell_{k+1}\rangle=0,\\
        &\,\langle A_{k+1}B_{k+1}\ell_k\rangle=\langle B_{k+1}A_{k+1}\ell_{k+2}\rangle=0,\\
        &\qquad\qquad\qquad\qquad\qquad\cdots
    \end{aligned}
\end{equation}
With medium- or short-rung insertions, we still have $R$ overlapping cycles, just longer cycles. Following the discussion around eq.(\ref{eq:cycle}), it is straightforward to show that the $R$ transfer matrices $M^{(r)}$ (see eq.(\ref{eq:shortcycle1}) and eq.(\ref{eq:shortcycle2})) will be different, but they still enjoy the key property that $M^{(r)}$ is proportional to $(\ell_{k_r+1})$, where the $r$-th long rung is between the $k_r$- and the $(k_r+1)$-th loops. The arguments in this subsection go through with little change. In the end, we would still obtain an $(R{+}1)$-variate polynomial $\mathcal D_{R+1}$ that is quadratic with respect to each variable. In other words, medium- or short-rung insertions do not change the rigidity of the LS, which is still an $R$-form.




\section{Discussions and outlook}
In this note, we have analyzed leading singularities of traintrack integrals and their degenerations such as ``half traintrack" integrals; the latter are the most general case with conventional ($0$-form) LS, which we computed to all loops in terms of a compact determinant. For general $L$-loop traintracks, we have proved that they have the expected rigidity $L{-}1$ by showing rigorously that their LS are given by integrals of $(L{-}1)$-forms; we have also proved for the first time that their LS parametrize $(L{-}1)$-dim Calabi-Yau manifolds. Both a direct computation of Cauchy's residues using Feynman parameters and the Schubert geometric approach in twistor space give the same result. 

Our results have opened up several new avenues for future investigations. Now that we have obtained CY manifolds from the LS, it would be highly desirable to study these geometric structures underlying this family of integrals: can we specify certain integration contours and compute periods (including LS) on these CY manifolds? What can we learn about the topological and other important properties of these manifolds to all loops, and what do they teach us about these integrals? Such properties of course depend on the kinematics: we have seen that even the dimension of CY manifolds is reduced when we take soft limits, and it would be intriguing to understand such degenerations better. An important lesson we learn is that, given an integral, we have to consider not only its own LS but the LS of its subtopologies; for example, an $L$-loop half traintrack involves up to $\lfloor \frac{L{-}1}{2} \rfloor$-dim CY manifolds, and even for massless traintracks we must consider subtopologies with massive legs. It is fascinating that a collection of CY manifolds, which are closely related to each other, are needed for a single integral. Needless to say, it would be highly desirable to study other Feynman integrals involving (higher-dimensional) CY manifolds~\cite{Bourjaily:2018yfy,Bourjaily:2019hmc,Bonisch:2021yfw,Bourjaily:2022bwx,Duhr:2022pch,Pogel:2022ken} and explore the possible connections between these geometries. 

Eventually we would like to better understand the analytic structures of the integrals themselves, which would require tools more powerful tools than the symbol/coproducts of MPL and eMPL functions. The representation in appendix~\ref{app-t-integral} of traintracks as $(L{-}1)$-fold integrals over $t$-deformed $(2L{+}2)$-gon can serve as a good starting point, for full traintracks and less rigid traintracks alike.
However, at the moment we do not know how to proceed in generality since the very notion of ``symbol/coproducts" is not well-defined beyond the simplest cases of rigidity $0$ (MPL)~\cite{Goncharov2002GaloisSO,Goncharov:2010jf,Duhr:2011zq,Duhr:2012fh} and $1$ (elliptic MPL)~\cite{Broedel:2018iwv,Kristensson:2021ani,Wilhelm:2022wow}. In the same spirit of this note, we propose to ask some really basic questions (similar to what we have done for LS): {\it e.g.}, which square roots, elliptic curves or CY geometries can appear for traintrack integrals? What we have learnt about LS forms and especially the underlying Schubert problems will be very useful in answering these questions, and eventually we hope to use the knowledge of ``symbology" to bootstrap these traintrack integrals, as have been done in~\cite{Morales:2022csr}. We end our discussion with  comments and speculations in this regard. 

\subsection{Comments on the ``symbology" of (degenerate) traintrack integrals}
Inspired by remarkable mathematical structures such as cluster algebras for a large class of Feynman integrals evaluating to MPL functions (see~\cite{Chicherin:2020umh,  He:2021esx, He:2021non, He:2021eec} and references therein), a new method called ``Schubert analysis" has been exploited to generate and explain the alphabets of such integrals~\cite{Yang:2022gko,He:2022ctv,He:2022tph}. The method has proved to be very powerful even for generating the alphabet of the eMPL symbols for the massive double-box ($L=2$ full traintrack)~\cite{Morales:2022csr}. Roughly speaking, the symbol letters for MPL ($\log$ of algebraic functions of kinematics) can be obtained by integrating the subLS $d\log$ 1-form along line intervals derived from cuts in twistor space, and for eMPL ``letters" the $d\log$ form is replaced by the elliptic form $\frac {dx}{y}$. We expect that similar analysis can be used at least for all (degenerate) traintracks which evaluate to MPL and eMPL functions, and with suitable generalizations also to higher rigidity cases, {\it e.g.}, by integrating higher forms with some higher-dimensional contour given by cuts in twistor space, which we leave to future works. 

Although it is unclear what ``symbology" means precisely for higher rigidity cases, here we simply consider iteratively taking their derivatives. Recall that the full traintrack is given by $(L{-}1)$-fold integrations of weight $L{+}1$ MPL functions, thus we expect that the rigidity-$L{-}1$ part, which involves $(L{-}1)$-dim CY manifold, appears in the second part of the coproduct $(L{+}1, L{-}1)$ in the sense of derivatives; the first part (weight $L{+}1$ function) should involve only functions with lower rigidity. For example, the weight-3 part of the $(3,1)$-coproduct of the $L=2$ full traintrack (double-box with 12 points in the massive case or 10 points in the massless case) are MPL functions~\cite{Morales:2022csr}; for $L=3$, we conjecture that its $(4,2)$-coproduct has weight-2 ``K3 functions" in the second part, and up to elliptic MPL functions in the first part. On the other hand, as we have seen in all known cases, the first two entries (or the first part of $(2,2L{-}2)$-coproducts) of such integrals must be MPL functions, and more specifically we conjecture them to be either the $2L{+}2 \choose 4$ box functions, or of the form $\log u \log v$ with $u,v$ being cross-ratios, which essentially follows from first-entry conditions and Steinmann relations~\cite{Gaiotto:2011dt,Caron-Huot:2016owq,Caron-Huot:2020bkp} (see~\cite{He:2021mme} for details). Higher-rigidity objects start to appear in subsequent ``entries". 

The symbol letters and higher-rigidity generalizations in general depend on LS of all subtopologies for a given Feynman integral: for rigidity $0$ these correspond to letters with square roots, and for non-zero rigidities they are periods over CY manifolds such as elliptic integrals and K3 integrals. For MPL/eMPL cases, the dependence can be made more concrete: these are non-rational letters of the form $\log \frac{a+\sqrt{\Delta}}{a+\sqrt{\Delta}}$,
and {\it elliptic} letters which changes sign under the flip $\omega_1 \to -\omega_1$ with $\omega_1$ given by an integral over the elliptic curve. Moreover, since the integral must be invariant under the Galois group which flips the sign of such periods, we conclude that the symbol is {\it even} under the flip of LS of any subtopology and {\it odd} under the flip of its own LS. This provides constraints on the symbol: for each term, odd letters with LS of a subtopology appear {\it even} times, while those with LS appear {\it odd} times; similar constraints apply for elliptic letters as well. 

Let us present some low-loop examples. The symbol of $L=2$ integral contains various odd letters which depend on one-loop square roots, and each odd letter indeed appears $0$ or $2$ times in the weight $4$ symbol, which can only appear in the second and third entries; the elliptic letters, which are odd under $\omega_1 \to -\omega_1$, must appear exclusively in the last entry (thus only once). Note that if we take the degeneration to half traintracks ({\it e.g.}, $b_1=b_2$), then the elliptic letters become odd letters in $\Delta_2$ which also only appear in the last entry (see~\cite{He:2022ujv, Wilhelm:2022wow} for details).

For $L=3$, we consider the $(4,2)$-coproduct of the full traintrack with rigidity 2: the second part involves some (unknown) weight-2 K3 functions, while the first part involves 0 or 2 appearances of odd letters with square roots from one- and two-loop subtopologies, as well as 0 or 2 appearances of elliptic letters with curves from two-loop subtopologies. 

\begin{figure}[H]
\centering
	\includegraphics[width=0.5\textwidth]{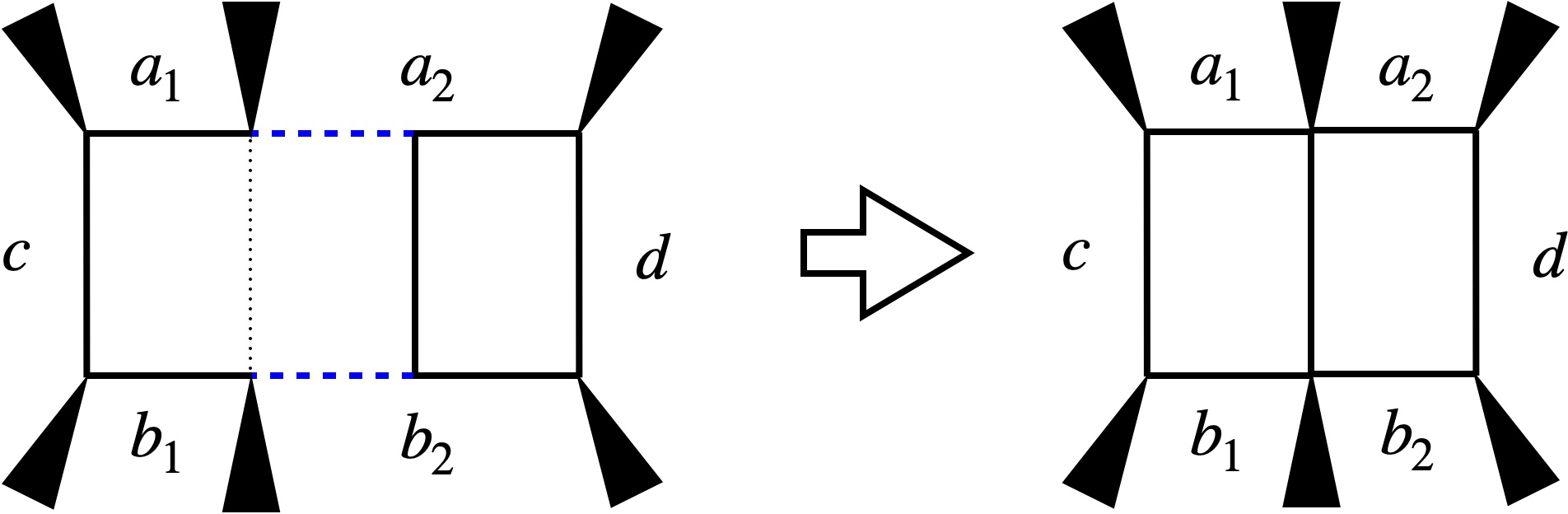}
 \caption{Elliptic subtopology of integral A (thin dotted line indicates deletion, blue dashed line indicates shrinkage)}
	\label{fig:integralA}
\end{figure}

Let us move to elliptic degenerations (rigidity $1$) of the $L=3$ traintrack; there are three inequivalent cases, depending on 6 dual points, obtained by taking two soft limits: one with elliptic (1-form) LS $A=\left[c\ud{a_1}{b_1}\ud{a_2}{b_2}\ud{a_2}{b_2}d\right]$, and two with 0-form LS, $B=\left[c\ud{a_1}{b_1}\ud{a_2}{b_1}\ud{a_2}{b_3}d\right]$ (alternating) and $C=\left[c\ud{a_1}{b_1}\ud{a_2}{b_1}\ud{a_3}{b_1}d\right]$ (all-up). Although all of them are elliptic MPL functions, their elliptic symbols have very different structures. For integral $A$, there must be 1 or 3 appearances of elliptic letters (odd under $\omega_1 \to -\omega_1$) in the symbol since the elliptic curve is from its own LS; we conjecture them to appear in $4,5,6$-entries. Note that it has a two-loop elliptic subtopology $\left[c\ud{a_1}{b_1 }\ud{a_2}{b_2}d\right]$ with the same LS (Fig.~\ref{fig:integralA}), thus $\omega_1$ itself can appear as an even letter (just like the square root itself in MPL cases). In principle, one could consider all-loop ladders which also evaluate to eMPL functions with the same elliptic curve, and again we expect odd letters to appear odd times (in entries 4 to $2L$) and the even letter $\omega_1$ itself to appear. 

\begin{figure}[H]
\centering
	\includegraphics[width=0.5\textwidth]{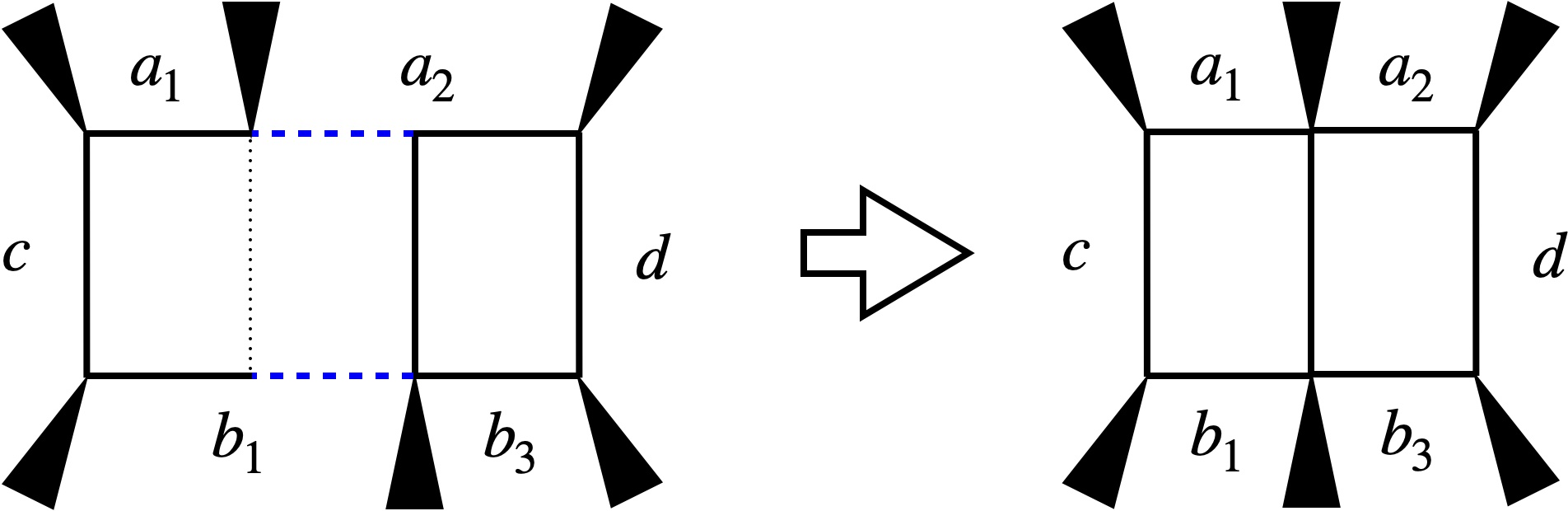}
 \caption{Elliptic subtopology of integral B}
	\label{fig:integralB}
\end{figure}

\begin{figure}[H]
\centering
	\includegraphics[width=0.5\textwidth]{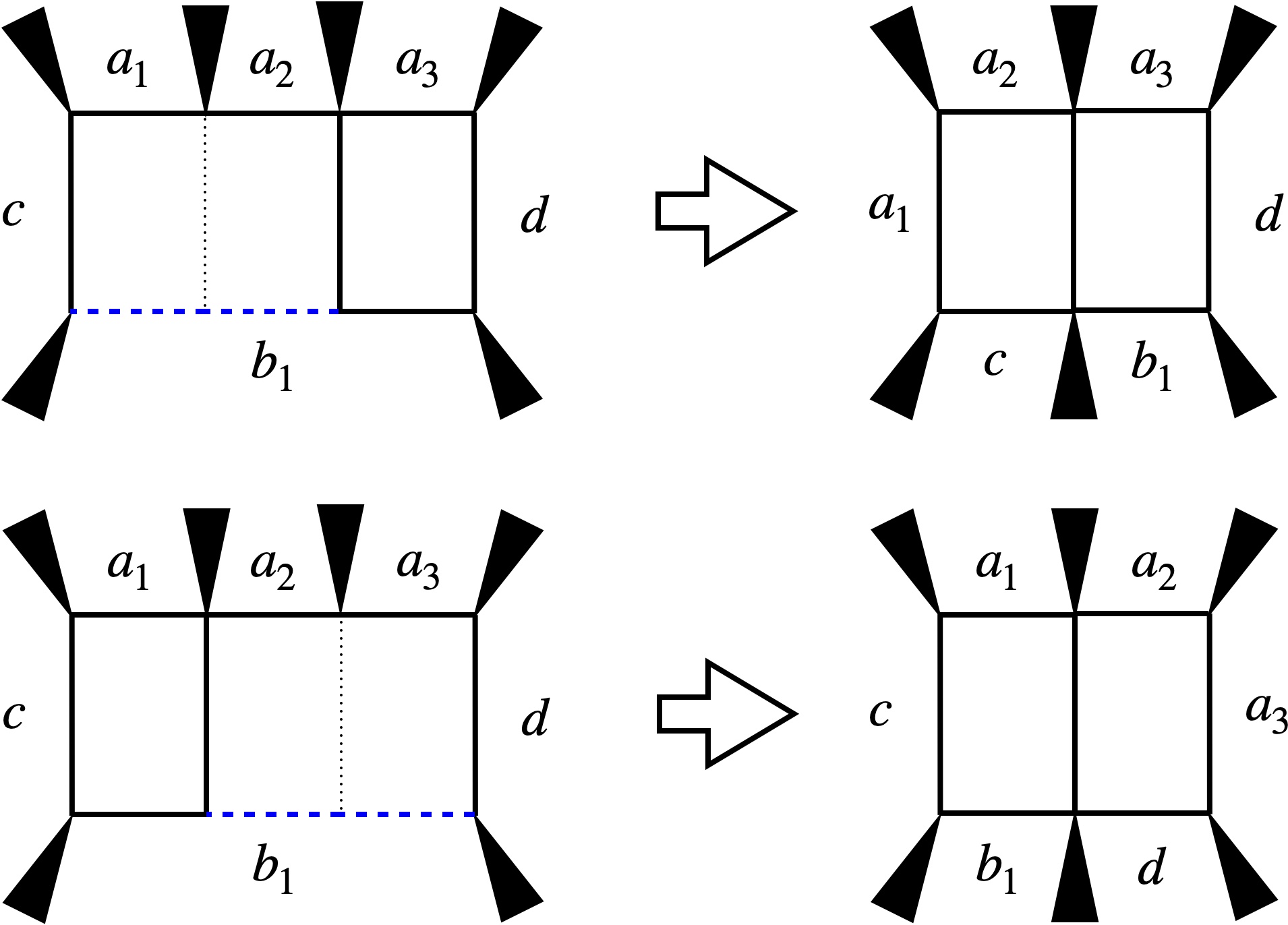}
 \caption{Elliptic subtopologies of integral C}
	\label{fig:integralC}
\end{figure}

For integral $B$, its rigidity is from the two-loop elliptic subtopology $\left[c\ud{a_1}{b_1}\ud{a_2}{b_3}d\right]$ (Fig.~\ref{fig:integralB}), thus the elliptic letters which are odd in $\omega_1 \to -\omega_1$ must appear $0$ or $2$ times (in entries $4,5,6$). For integral $C$, interestingly its rigidity is from two different two-loop elliptic subtopologies $\left[c\ud{a_1}{b_1}\ud{a_2}{d}a_{3}\right]$ and $\left[a_{1}\ud{a_2}{c}\ud{a_3}{b_{1}}d\right]$ (Fig.~\ref{fig:integralC}), thus elliptic letters (including those that mix the two elliptic curves), which are odd with respect to either one of the curves, must appear $0$ or $2$ times (in entries $4,5,6$). In addition, both $B$ and $C$ have $0$-form LS, thus only in the last entry do we have odd letters involving the LS square root (which can also mix with elliptic ones); odd letters with respect to other square roots (LS of one- or two-loop subtopologies) must appear even times. 

Although we can confirm that they are indeed elliptic MPLs, we have not been able to directly compute (the symbol) of these integrals, which we leave to future works. However, one can compute/bootstrap further degenerations which evaluate to MPL functions, {\it e.g.}, all three integrals can degenerate to $\left[c\ud{a_1}{b_1}\ud{a_2}{b_1}\ud{a_2}{b_1}d\right]$, which depends on $5$ dual points. Using Schubert analysis, we have obtained an alphabet with $42$ letters and bootstrapped its symbol\footnote{We thank Qinglin Yang for this computation.}, which serves as ``boundary data" as well as evidence supporting our claims regarding integrals $A,B$ and $C$.


\section*{Acknowledgement} We thank Zhenjie Li,  Ellis Ye Yuan, Chi Zhang, and especially Qinglin Yang for stimulating discussions and collaborations on related projects. This research is supported in part by National Natural Science Foundation of China under Grant No. 11935013, 11947301, 12047502, 12047503, 12225510.

{\bf Note added}: While this work was in progress, we were made aware of a related upcoming work \cite{McLeod:2023qdf}, which studies all-loop elliptic ladder integrals. 

\appendix
\section{Details for the residue from Feynman parameters}\label{app-details-feynman}
In this appendix, we give more details on the derivation from $\Delta_{L}^{(2)}$ to $\Delta_{L}^{(3)}$, including our method of taking composite residues. For definiteness, we consider an even-loop alternating traintrack, so that $b_{L-1}=b_L$ and $a_{L-2}=a_{L-1}$. The method applies to all half traintracks. Following section~\ref{sec:fey-ls},
\begin{align}\label{eq:delta2}
    \Delta_L^{(2)}&=\left[\x{b_{L}}{d}\det(a_{L-1},R_{L-2};a_{L},b_{L})-\x{a_{L}}{b_{L}}\det(a_{L-1},R_{L-2};d,b_{L})\right]^2\nonumber\\
    &-4\x{a_{L}}{d}\x{b_{L}}{d}\x{a_{L}}{b_{L}}\x{R_{L-2}}{a_{L-1}}\x{b_{L}}{R_{L-2}}\x{b_{L}}{a_{L-1}}\nonumber\\
    &\equiv\left[A_{L}^{(2)}\right]^{2}-4B_{L}^{(2)}.
\end{align}

In order to take the composite residue at $(R_{L-2},R_{L-2})=0$, we split the definition eq.(\ref{eq-gl1}) of $(R_{L-2})$ with $\gamma_{L-2}=1$ into two steps:
\begin{equation}
    (R_{L-2})=(\tilde R_{L-2})+\alpha_{L-2}(a_{L-2}),\quad(\tilde R_{L-2})\equiv(R_{L-3})+\beta_{L-2}(b_{L-2}),
\end{equation}
and first take the residue with respect to $\alpha_{L-2}$. Using $a_{L-2}=a_{L-1}$, we have
\begin{equation}
    \alpha_{L-2}=-\frac{(\tilde R_{L-2},\tilde R_{L-2})}{2(a_{L-1},\tilde R_{L-2})}\equiv\alpha_{L-2}^*.
\end{equation}
Plugging this into eq.(\ref{eq:delta2}),
\begin{align}
    A_L^{(2)}\xlongequal{\alpha_{L-2}=\alpha_{L-2}^*}&(b_L,d)\det(a_{L-1},\tilde R_{L-2};a_L,b_L)-(a_L,b_L)\det(a_{L-1},\tilde R_{L-2};d,b_L),\\
    B_L^{(2)}\xlongequal{\alpha_{L-2}=\alpha_{L-2}^*}&(a_L,d)(b_L,d)(a_L,b_L)(a_{L-1},b_L)\nonumber\\
    &\times\left[(\tilde R_{L-2},a_{L-1})(b_L,\tilde R_{L-2})-\frac12(\tilde R_{L-2},\tilde R_{L-2})(a_{L-1},b_L)\right].
\end{align}

Next, take the residue of the Jacobian $(a_{L-1},\tilde R_{L-2})=0$ with respect to $\beta_{L-2}$:
\begin{equation}
    \beta_{L-2}=-\frac{(a_{L-1},R_{L-3})}{(a_{L-1},b_{L-2})}\equiv\beta_{L-2}^*.
\end{equation}
Moving the new Jacobian factor $(a_{L-1},b_{L-2})$ into the square root, we obtain (on the support of $(R_{L-3},R_{L-3})=0$)
\begin{gather}
    \mathop{\rm compRes}_{(R_{L-2},R_{L-2})=0}\frac1{(R_{L-2},R_{L-2})\sqrt{\Delta_L^{(2)}}}=\frac1{\sqrt{\Delta_L^{(3)}}},\\
    \Delta_L^{(3)}\equiv(a_{L-1},b_{L-2})^2\Delta_L^{(2)}\Big|_{\substack{\alpha_{L-2}=\alpha_{L-2}^*\\\beta_{L-2}=\beta_{L-2}^*}}=\left[A_L^{(3)}\right]^2-4B_L^{(3)},
\end{gather}
where 
\begin{align}
 	A_L^{(3)}&=\det\begin{pmatrix}
 		\x{d}{b_{L}}&\x{a_{L}}{b_{L}}&0\\ \x{d}{a_{L-1}}&\x{a_{L}}{a_{L-1}}&\x{b_{L}}{a_{L-1}}\\ \det(R_{L-3},b_{L-2};d,a_{L-1})&\det(R_{L-3},b_{L-2};a_{L},a_{L-1})&\det(R_{L-3},b_{L-2};b_{L},a_{L-1})
 	\end{pmatrix}\nonumber\\
  &=\det\begin{pmatrix}
			\x{d}{b_{L}}&\x{a_{L}}{b_{L}}&0&0\\ \x{d}{a_{L-1}}&\x{a_{L}}{a_{L-1}}&\x{b_{L}}{a_{L-1}}&0\\\x{d}{b_{L-2}}&\x{a_{L}}{b_{L-2}}&\x{b_{L-2}}{b_{L}}&\x{b_{L-2}}{a_{L-1}}\\ \x{d}{R_{L-3}}&\x{a_{L}}{R_{L-3}}&\x{b_{L}}{R_{L-3}}&\x{a_{L-1}}{R_{L-3}} 
		\end{pmatrix},\label{detA-2}\\
  B_L^{(3)}&=B_{L}^{(2)}\x{a_{L-1}}{b_{L-2}}\x{b_{L}}{a_{L-1}}\Big|_{\substack{\x{R_{L-2}}{b_{L}}\to\x{R_{L-3}}{b_{L-2}}\\\x{R_{L-2}}{a_{L-1}}\to\x{R_{L-3}}{a_{L-1}}}}\,.
\end{align}

The matrix $\mathcal{A}_{L}^{(3)}$ such that $A_L^{(3)}=\det\mathcal{A}_{L}^{(3)}$ can be obtained from $\mathcal A_L^{(2)}$ in eq.(\ref{eq:amat2}) by
\begin{enumerate}
    \item Changing the row labels from $\{b_L,a_{L-1},R_{L-2}\}$ to $\{b_L,a_{L-1},b_{L-2},R_{L-3}\}$ and extending the column labels from $\{d,a_L,b_L\}$ to $\{d,a_L,b_L,a_{L-1}\}$;
    \item Filling in the blanks in the new row and the new column as if the matrix were a Gram matrix; and
    \item Setting to zero all but the last two rows of the new column.
\end{enumerate}

\section{Example: 1-form ``elliptic" leading singularities}\label{app-detail-ell}
\begin{figure}[h]
		\centering
	\includegraphics[width=0.45\textwidth]{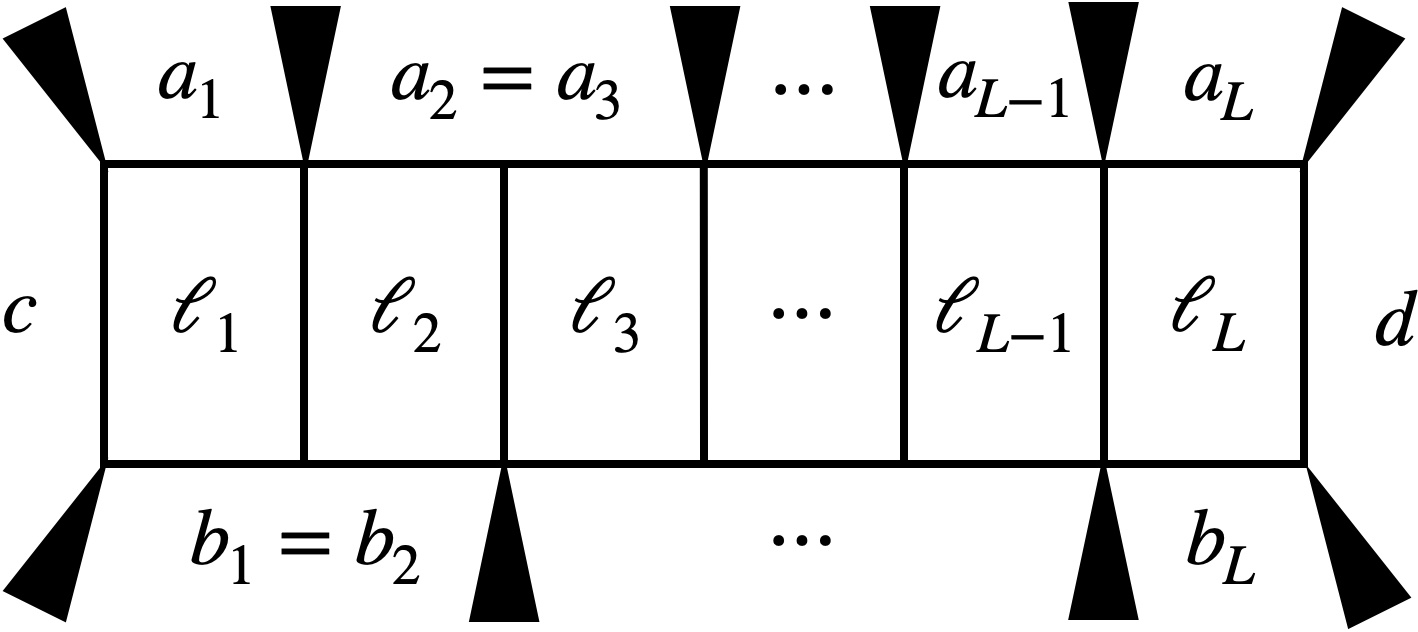}\qquad
 \includegraphics[width=0.45\textwidth]{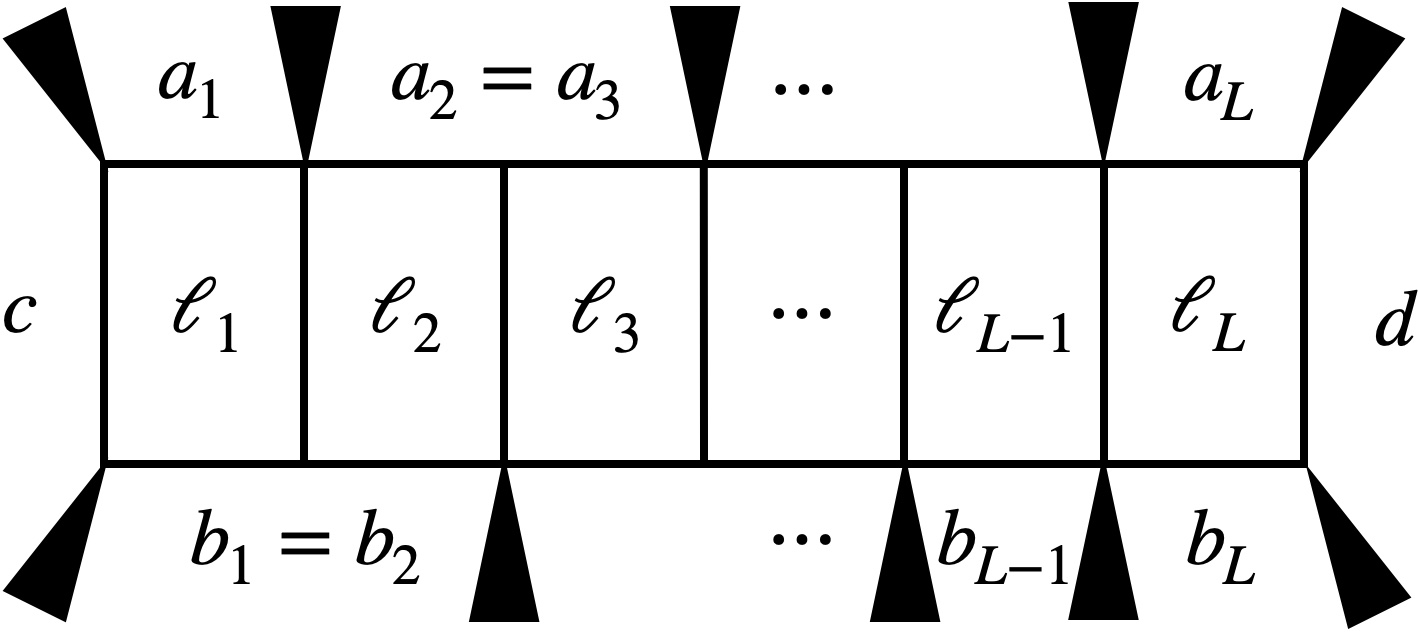}
	\caption{The degenerate traintracks with ``elliptic" LS for odd and even $L$}
	\label{fig:traintrack_ell}
\end{figure}

Consider a more complicate case as an example, when there is an ``elliptic" (1-form) LS, which has been studied in~\cite{Bourjaily:2020hjv,Bourjaily:2021vyj}. In general, 1-form LS arises whenever there is exactly one long rung.
In this appendix, we consider the ``almost-half traintracks'' in Fig.~\ref{fig:traintrack_ell}. Our method applies with little change to cases where the long rung appears at different locations.


When $L=2$, the ``elliptic" leading singularity is 
\begin{equation}
	{\bf LS}^{(2)}_{\rm elliptic}={\cal N}_2 \int \frac{ d\alpha}{\sqrt{\mathcal{Q}_{2}(\alpha)}}\,,
\end{equation} 
where $\mathcal{Q}_{2}(\alpha)$ can be expanded as the quartic polynomial of $\alpha\equiv\alpha_{1}$ ,
\begin{equation}\label{ls-e}
	\begin{aligned}
		\mathcal{Q}_{2}(\alpha)=&c_{1}\alpha^{4} +c_{2}\alpha^{3}+c_{3}\alpha^{2}+c_{4}\alpha+ c_{5}\,,
	\end{aligned}
\end{equation}
For general $L$, we apply the rule~\eqref{eq:subs-half} recursively to~\eqref{ls-e}, which leads to
\begin{equation}
	{\bf LS}^{(L)}_{\rm elliptic}={\cal N}_L \int\frac{d\alpha}{\sqrt{\mathcal{Q}_{L}(\alpha)}}\,,
\end{equation} 
where the coefficients of $\mathcal{Q}_{L}(\alpha)$ get deformed. Specifically, 
\begin{equation}\label{ls-coe}
	\begin{aligned}
		c_{1}=&\x{a_{L-1}}{R_{L-2}}^{2}\Delta(a_{L},d,b_{L},a_{L-1})\\
		c_{2}=&2\x{a_{L-1}}{R_{L-2}}A(a_{L},d,b_{L},a_{L-1})C(R_{L-2},a_{L-1},a_{L},d,b_{L},b_{L-1})\\
		&-4\x{a_{L-1}}{R_{L-2}}\x{a_{L}}{d}\x{b_{L}}{d}(\x{b_{L}}{a_{L-1}}A(R_{L-2},a_{L-1},a_{L},b_{L-1})+(a_{L}\leftrightarrow b_{L}))\\
		c_{3}=&C(R_{L-2},a_{L-1},a_{L},d,b_{L},b_{L-1})^{2}\\
  &+2\x{a_{L-1}}{R_{L-2}}\x{R_{L-2}}{b_{L-1}}A(a_{L},d,b_{L},a_{L-1})A(a_{L},d,b_{L},b_{L-1})\\
		&-4\x{a_{L}}{d}\x{b_{L}}{d}(A(R_{L-2},a_{L-1},a_{L},b_{L-1})A(R_{L-2},a_{L-1},b_{L},b_{L-1})\\
		& +\x{a_{L-1}}{R_{L-2}}\x{b_{L}}{a_{L-1}}\x{R_{L-2}}{b_{L-1}}\x{a_{L}}{b_{L-1}}+(a_{L}\leftrightarrow b_{L}))\\
		c_{4}=&2\x{R_{L-2}}{b_{L-1}}A(a_{L},d,b_{L},b_{L-1})C(R_{L-2},a_{L-1},a_{L},d,b_{L},b_{L-1})\\
		&-4\x{R_{L-2}}{b_{L-1}}\x{a_{L}}{d}\x{b_{L}}{d}(A(R_{L-2},a_{L-1},b_{L},b_{L-1})\x{a_{L}}{b_{L-1}}+(a_{L}\leftrightarrow b_{L}))\\
		c_{5}=&\x{R_{L-2}}{b_{L-1}}^{2}\Delta(a_{L},d,b_{L},b_{L-1})
	\end{aligned}
\end{equation}
where
\begin{align}
	A(a,b,c,d)&\equiv\x{a}{b}\x{c}{d}+\x{a}{d}\x{b}{c}-\x{a}{c}\x{b}{d}\,,\\
    \Delta(a,b,c,d)&\equiv A(a,b,c,d)^{2}-4\x{a}{b}\x{c}{d}\x{a}{d}\x{b}{c}\,,\\
    C(a,b,c,d,e,f)&\equiv\x{a}{b}A(c,d,e,f)+\x{a}{f}A(c,d,e,b)-\x{b}{f}A(c,d,e,a)\,.
\end{align}
	
For $L=2$, we have $R_{L-2}=R_{0}\equiv c$. For $L>2$, $R_{L-2}$ is found by iteratively using the rule~\eqref{eq:subs-half}.

\section{$L$-loop traintracks from $t$-deformed $(2L+2)$-gons}\label{app-t-integral}
Here we present the explicit form of an $L$-loop (massive) traintrack integral as an $(L{-}1)$-fold integral of a $(2L{+}2)$-gon with deformed kinematics, by combining factors in the denominator using Feynman parameters. We start from the loop-by-loop Feynman-parametrized representation:
\eq{\traintrack\!=\!\int_0^{\infty}\!\!\proj{d^{2L+1}\!\vec{X}}
 \frac{\mathcal{N}_L}{\x{R_1}{R_1}\cdots
\x{R_{L-1}}{R_{L-1}}\x{R_L'}{R_L'}^{2}}\,,}
where the projective measure $\proj{d^{2L+1}\!\vec{X}}$ is over the $2L+2$ Feynman parameters $\{\alpha_{0},\alpha_1,\dots, \alpha_L,\beta_{0},\beta_{1},\dots,\beta_L\}$ in $\mathbb{CP}^{2L+1}$:
\begin{gather}
    (R_k)\!\equiv\!(R_{k-1})\pl\alpha_k(a_k)\pl\beta_k(b_k), \quad\text{for }k=1,\cdots,L,\\
(R_0)\!\equiv\! \alpha_0(c),\quad (R_L')\!\equiv\!(R_{L-1})\pl\alpha_L(a_L)\pl\beta_L(b_L)\pl\beta_{0}(d).
\end{gather}
Introduce new parameters to combine the denominators,
\begin{multline}
    \frac{1}{\x{R_1}{R_1}\cdots\x{R_{L-1}}{R_{L-1}}
\x{R_L}{R_L}^{2}}\\
=\int_0^\infty\frac{\proj{d^{L-1}\!\vec{t}}\Gamma(L+1)t_{L}}{\left(t_{1}\x{R_1}{R_1}+\dots+t_{L-1}\x{R_{L-1}}{R_{L-1}}+t_{L}\x{R_L'}{R_L'}\right)^{L+1}}\,,
\end{multline}
where the projective measure $\proj{d^{L-1}\!\vec{t}}$ is over the $L$ new Feynman parameters $\{t_{1},\dots,t_L\}$ in $\mathbb{CP}^{L-1}$. Thus, we obtain the key result:
\begin{equation}\label{eq-t-integral}
\begin{aligned}
\traintrack\!&=\!\int_0^{\infty}\!\!\proj{d^{2L+1}\!\vec{X}} \proj{d^{L-1}\!\vec{t}}
 \frac{\mathcal{N}_L\Gamma(L+1)t_{L}}{\left(t_{1}\x{R_1}{R_1}+\dots+t_{L-1}\x{R_{L-1}}{R_{L-1}}+t_{L}\x{R_L'}{R_L'}\right)^{L+1}}\\
 &=\int_0^\infty\proj{d^{L-1}\!\vec{t}} \!\int_0^{\infty}\!\!\proj{d^{2L+1}\!\vec{X}} 
 \frac{\mathcal{N}_L\Gamma(L+1)t_{L}}{\left(\vec{X}.\mathbb{Q}(\vec{t}).\vec{X}\right)^{L+1}}\,,
\end{aligned}
\end{equation}
where the quadric $\vec{X}.\mathbb{Q}(\vec{t}).\vec{X}$ can be viewed as the one-loop $(2L{+}2)$-gon integral deformed by the $t$-parameters. When $[t_1:\cdots:t_{L-1}:t_L]=[0:\cdots:0:1]$, the quadric reduces to that of the $(2L{+}2)$-gon~\cite{Arkani-Hamed:2017ahv}.

The precise expression of $\mathbb Q(\vec t)$ is easily obtain from the undeformed $(2L{+}2)$-gon quadric $\vec{X}.\mathbb{Q}.\vec{X}$ by deforming
\begin{equation}
    (i,j)\to(i,j)\times\left(t_L+\sum_{\substack{k=1\\(R_k,R_k)\supset(i,j)}}^{L-1}t_k\right).
\end{equation}

For example, an $L=2$ traintrack integral can be seen as the integral of a $t$-deformed hexagon:
\begin{equation}
\begin{aligned}
\mathfrak T^{(2)}&=\int_0^\infty \proj{d^{1}\!\vec{t}} \!\int_0^{\infty}\!\!\proj{d^{5}\!\vec{X}} 
 \frac{2\mathcal{N}_2\,t_{2}}{\left(\vec{X}.\mathbb{Q}(\vec{t}).\vec{X}\right)^{3}}\,,
\end{aligned}
\end{equation}
where $\vec{X}=(\alpha_{0},\alpha_{1},\alpha_{2},\beta_{0},\beta_{1},\beta_{2})$ and $\vec t=(t_1,t_2)$. The undeformed quadric $\mathbb Q$ is simply a Gram matrix:
\begin{equation}
    \mathbb{Q}=\begin{pmatrix}
      0&\x{c}{a_{1}}&\x{c}{a_{2}}&\x{c}{d}&\x{c}{b_{1}}&\x{c}{b_{2}}\\
      \x{c}{a_{1}}&0&\x{a_{1}}{a_{2}}&\x{a_{1}}{d}&\x{a_{1}}{b_{1}}&\x{a_{1}}{b_{2}}\\
      \x{c}{a_{2}}&\x{a_{1}}{a_{2}}&0&\x{a_{2}}{d}&\x{a_{2}}{b_{1}}&\x{a_{2}}{b_{2}}\\
      \x{c}{d}&\x{a_{1}}{d}&\x{a_{2}}{d}&0&\x{b_{1}}{d}&\x{b_{2}}{d}\\
      \x{c}{b_{1}}&\x{a_{1}}{b_{1}}&\x{a_{2}}{b_{1}}&\x{d}{b_{1}}&0&\x{b_{1}}{b_{2}}\\
      \x{c}{b_{2}}&\x{a_{1}}{b_{2}}&\x{a_{2}}{b_{2}}&\x{d}{b_{2}}&\x{b_{1}}{b_{2}}&0
    \end{pmatrix}.
\end{equation}
Since $(R_1)=\alpha_0(c)+\alpha_1(a_1)+\beta_1(b_1)$, the factor $(R_1,R_1)$ contains $(c,a_1)$, $(c,b_1)$, and $(a_1,b_1)$. Hence,
\begin{equation}
    \mathbb{Q}(\vec{t})=\begin{pmatrix}
      0&(t_{1}+t_{2})\x{c}{a_{1}}&t_{2}\x{c}{a_{2}}&t_{2}\x{c}{d}&(t_{1}+t_{2})\x{c}{b_{1}}&t_{2}\x{c}{b_{2}}\\
      (t_{1}+t_{2})\x{c}{a_{1}}&0&t_{2}\x{a_{1}}{a_{2}}&t_{2}\x{a_{1}}{d}&(t_{1}+t_{2})\x{a_{1}}{b_{1}}&t_{2}\x{a_{1}}{b_{2}}\\
      t_{2}\x{c}{a_{2}}&t_{2}\x{a_{1}}{a_{2}}&0&t_{2}\x{a_{2}}{d}&t_{2}\x{a_{2}}{b_{1}}&t_{2}\x{a_{2}}{b_{2}}\\
      t_{2}\x{c}{d}&t_{2}\x{a_{1}}{d}&t_{2}\x{a_{2}}{d}&0&t_{2}\x{b_{1}}{d}&t_{2}\x{b_{2}}{d}\\
      (t_{1}+t_{2})\x{c}{b_{1}}&(t_1+t_{2})\x{a_{1}}{b_{1}}&t_{2}\x{a_{2}}{b_{1}}&t_{2}\x{d}{b_{1}}&0&t_{2}\x{b_{1}}{b_{2}}\\
      t_{2}\x{c}{b_{2}}&t_{2}\x{a_{1}}{b_{2}}&t_{2}\x{a_{2}}{b_{2}}&t_{2}\x{d}{b_{2}}&t_{2}\x{b_{1}}{b_{2}}&0
    \end{pmatrix}.
\end{equation}

	\bibliographystyle{JHEP}
\bibliography{reference}

\providecommand{\href}[2]{#2}\begingroup\raggedright\begin{thebibliography}{10}

\bibitem{Chen:1977oja}
K.-T. Chen, {\it {Iterated path integrals}},  {\em Bull. Am. Math. Soc.} {\bf
  83} (1977) 831--879.

\bibitem{Goncharov1995GeometryOC}
A.~B. Goncharov, {\it Geometry of configurations, polylogarithms, and motivic
  cohomology},  {\em Advances in Mathematics} {\bf 114} (1995) 197--318.

\bibitem{Goncharov:1998kja}
A.~B. Goncharov, {\it {Multiple polylogarithms, cyclotomy and modular
  complexes}},  {\em Math. Res. Lett.} {\bf 5} (1998) 497--516,
  [\href{http://arxiv.org/abs/1105.2076}{{\tt arXiv:1105.2076}}].

\bibitem{Remiddi:1999ew}
E.~Remiddi and J.~A.~M. Vermaseren, {\it {Harmonic polylogarithms}},  {\em Int.
  J. Mod. Phys. A} {\bf 15} (2000) 725--754,
  [\href{http://arxiv.org/abs/hep-ph/9905237}{{\tt hep-ph/9905237}}].

\bibitem{Borwein:1999js}
J.~M. Borwein, D.~M. Bradley, D.~J. Broadhurst, and P.~Lisonek, {\it {Special
  values of multiple polylogarithms}},  {\em Trans. Am. Math. Soc.} {\bf 353}
  (2001) 907--941, [\href{http://arxiv.org/abs/math/9910045}{{\tt
  math/9910045}}].

\bibitem{Moch:2001zr}
S.~Moch, P.~Uwer, and S.~Weinzierl, {\it {Nested sums, expansion of
  transcendental functions and multiscale multiloop integrals}},  {\em J. Math.
  Phys.} {\bf 43} (2002) 3363--3386,
  [\href{http://arxiv.org/abs/hep-ph/0110083}{{\tt hep-ph/0110083}}].

\bibitem{Goncharov2002GaloisSO}
A.~B. Goncharov, {\it Galois symmetries of fundamental groupoids and
  noncommutative geometry},  {\em Duke Mathematical Journal} {\bf 128} (2002)
  209--284.

\bibitem{Goncharov:2010jf}
A.~B. Goncharov, M.~Spradlin, C.~Vergu, and A.~Volovich, {\it {Classical
  Polylogarithms for Amplitudes and Wilson Loops}},  {\em Phys. Rev. Lett.}
  {\bf 105} (2010) 151605, [\href{http://arxiv.org/abs/1006.5703}{{\tt
  arXiv:1006.5703}}].

\bibitem{Duhr:2011zq}
C.~Duhr, H.~Gangl, and J.~R. Rhodes, {\it {From polygons and symbols to
  polylogarithmic functions}},  {\em JHEP} {\bf 10} (2012) 075,
  [\href{http://arxiv.org/abs/1110.0458}{{\tt arXiv:1110.0458}}].

\bibitem{Duhr:2012fh}
C.~Duhr, {\it {Hopf algebras, coproducts and symbols: an application to Higgs
  boson amplitudes}},  {\em JHEP} {\bf 08} (2012) 043,
  [\href{http://arxiv.org/abs/1203.0454}{{\tt arXiv:1203.0454}}].

\bibitem{Laporta:2004rb}
S.~Laporta and E.~Remiddi, {\it {Analytic treatment of the two loop equal mass
  sunrise graph}},  {\em Nucl. Phys. B} {\bf 704} (2005) 349--386,
  [\href{http://arxiv.org/abs/hep-ph/0406160}{{\tt hep-ph/0406160}}].

\bibitem{Muller-Stach:2012tgz}
S.~Muller-Stach, S.~Weinzierl, and R.~Zayadeh, {\it {From motives to
  differential equations for loop integrals}},  {\em PoS} {\bf LL2012} (2012)
  005, [\href{http://arxiv.org/abs/1209.3714}{{\tt arXiv:1209.3714}}].

\bibitem{Brown2011MultipleEP}
F.~Brown and A.~Levin, {\it Multiple elliptic polylogarithms},  {\em arXiv:
  Number Theory} (2011).

\bibitem{Bloch:2013tra}
S.~Bloch and P.~Vanhove, {\it {The elliptic dilogarithm for the sunset graph}},
   {\em J. Number Theor.} {\bf 148} (2015) 328--364,
  [\href{http://arxiv.org/abs/1309.5865}{{\tt arXiv:1309.5865}}].

\bibitem{Adams:2013nia}
L.~Adams, C.~Bogner, and S.~Weinzierl, {\it {The two-loop sunrise graph with
  arbitrary masses}},  {\em J. Math. Phys.} {\bf 54} (2013) 052303,
  [\href{http://arxiv.org/abs/1302.7004}{{\tt arXiv:1302.7004}}].

\bibitem{Adams:2014vja}
L.~Adams, C.~Bogner, and S.~Weinzierl, {\it {The two-loop sunrise graph in two
  space-time dimensions with arbitrary masses in terms of elliptic
  dilogarithms}},  {\em J. Math. Phys.} {\bf 55} (2014), no.~10 102301,
  [\href{http://arxiv.org/abs/1405.5640}{{\tt arXiv:1405.5640}}].

\bibitem{Adams:2015gva}
L.~Adams, C.~Bogner, and S.~Weinzierl, {\it {The two-loop sunrise integral
  around four space-time dimensions and generalisations of the Clausen and
  Glaisher functions towards the elliptic case}},  {\em J. Math. Phys.} {\bf
  56} (2015), no.~7 072303, [\href{http://arxiv.org/abs/1504.03255}{{\tt
  arXiv:1504.03255}}].

\bibitem{Adams:2015ydq}
L.~Adams, C.~Bogner, and S.~Weinzierl, {\it {The iterated structure of the
  all-order result for the two-loop sunrise integral}},  {\em J. Math. Phys.}
  {\bf 57} (2016), no.~3 032304, [\href{http://arxiv.org/abs/1512.05630}{{\tt
  arXiv:1512.05630}}].

\bibitem{Adams:2016xah}
L.~Adams, C.~Bogner, A.~Schweitzer, and S.~Weinzierl, {\it {The kite integral
  to all orders in terms of elliptic polylogarithms}},  {\em J. Math. Phys.}
  {\bf 57} (2016), no.~12 122302, [\href{http://arxiv.org/abs/1607.01571}{{\tt
  arXiv:1607.01571}}].

\bibitem{Adams:2017ejb}
L.~Adams and S.~Weinzierl, {\it {Feynman integrals and iterated integrals of
  modular forms}},  {\em Commun. Num. Theor. Phys.} {\bf 12} (2018) 193--251,
  [\href{http://arxiv.org/abs/1704.08895}{{\tt arXiv:1704.08895}}].

\bibitem{Adams:2017tga}
L.~Adams, E.~Chaubey, and S.~Weinzierl, {\it {Simplifying Differential
  Equations for Multiscale Feynman Integrals beyond Multiple Polylogarithms}},
  {\em Phys. Rev. Lett.} {\bf 118} (2017), no.~14 141602,
  [\href{http://arxiv.org/abs/1702.04279}{{\tt arXiv:1702.04279}}].

\bibitem{Bogner:2017vim}
C.~Bogner, A.~Schweitzer, and S.~Weinzierl, {\it {Analytic continuation and
  numerical evaluation of the kite integral and the equal mass sunrise
  integral}},  {\em Nucl. Phys. B} {\bf 922} (2017) 528--550,
  [\href{http://arxiv.org/abs/1705.08952}{{\tt arXiv:1705.08952}}].

\bibitem{Broedel:2017kkb}
J.~Broedel, C.~Duhr, F.~Dulat, and L.~Tancredi, {\it {Elliptic polylogarithms
  and iterated integrals on elliptic curves. Part I: general formalism}},  {\em
  JHEP} {\bf 05} (2018) 093, [\href{http://arxiv.org/abs/1712.07089}{{\tt
  arXiv:1712.07089}}].

\bibitem{Broedel:2017siw}
J.~Broedel, C.~Duhr, F.~Dulat, and L.~Tancredi, {\it {Elliptic polylogarithms
  and iterated integrals on elliptic curves II: an application to the sunrise
  integral}},  {\em Phys. Rev. D} {\bf 97} (2018), no.~11 116009,
  [\href{http://arxiv.org/abs/1712.07095}{{\tt arXiv:1712.07095}}].

\bibitem{Adams:2018yfj}
L.~Adams and S.~Weinzierl, {\it {The $\varepsilon$-form of the differential
  equations for Feynman integrals in the elliptic case}},  {\em Phys. Lett. B}
  {\bf 781} (2018) 270--278, [\href{http://arxiv.org/abs/1802.05020}{{\tt
  arXiv:1802.05020}}].

\bibitem{Broedel:2018iwv}
J.~Broedel, C.~Duhr, F.~Dulat, B.~Penante, and L.~Tancredi, {\it {Elliptic
  symbol calculus: from elliptic polylogarithms to iterated integrals of
  Eisenstein series}},  {\em JHEP} {\bf 08} (2018) 014,
  [\href{http://arxiv.org/abs/1803.10256}{{\tt arXiv:1803.10256}}].

\bibitem{Broedel:2018qkq}
J.~Broedel, C.~Duhr, F.~Dulat, B.~Penante, and L.~Tancredi, {\it {Elliptic
  Feynman integrals and pure functions}},  {\em JHEP} {\bf 01} (2019) 023,
  [\href{http://arxiv.org/abs/1809.10698}{{\tt arXiv:1809.10698}}].

\bibitem{Honemann:2018mrb}
I.~H\"onemann, K.~Tempest, and S.~Weinzierl, {\it {Electron self-energy in QED
  at two loops revisited}},  {\em Phys. Rev. D} {\bf 98} (2018), no.~11 113008,
  [\href{http://arxiv.org/abs/1811.09308}{{\tt arXiv:1811.09308}}].

\bibitem{Bogner:2019lfa}
C.~Bogner, S.~M\"uller-Stach, and S.~Weinzierl, {\it {The unequal mass sunrise
  integral expressed through iterated integrals on $\overline{\mathcal
  M}_{1,3}$}},  {\em Nucl. Phys. B} {\bf 954} (2020) 114991,
  [\href{http://arxiv.org/abs/1907.01251}{{\tt arXiv:1907.01251}}].

\bibitem{Broedel:2019hyg}
J.~Broedel, C.~Duhr, F.~Dulat, B.~Penante, and L.~Tancredi, {\it {Elliptic
  polylogarithms and Feynman parameter integrals}},  {\em JHEP} {\bf 05} (2019)
  120, [\href{http://arxiv.org/abs/1902.09971}{{\tt arXiv:1902.09971}}].

\bibitem{Duhr:2019rrs}
C.~Duhr and L.~Tancredi, {\it {Algorithms and tools for iterated Eisenstein
  integrals}},  {\em JHEP} {\bf 02} (2020) 105,
  [\href{http://arxiv.org/abs/1912.00077}{{\tt arXiv:1912.00077}}].

\bibitem{Walden:2020odh}
M.~Walden and S.~Weinzierl, {\it {Numerical evaluation of iterated integrals
  related to elliptic Feynman integrals}},  {\em Comput. Phys. Commun.} {\bf
  265} (2021) 108020, [\href{http://arxiv.org/abs/2010.05271}{{\tt
  arXiv:2010.05271}}].

\bibitem{Weinzierl:2020fyx}
S.~Weinzierl, {\it {Modular transformations of elliptic Feynman integrals}},
  {\em Nucl. Phys. B} {\bf 964} (2021) 115309,
  [\href{http://arxiv.org/abs/2011.07311}{{\tt arXiv:2011.07311}}].

\bibitem{Giroux:2022wav}
M.~Giroux and A.~Pokraka, {\it {Loop-by-loop Differential Equations for Dual
  (Elliptic) Feynman Integrals}},  \href{http://arxiv.org/abs/2210.09898}{{\tt
  arXiv:2210.09898}}.

\bibitem{Kristensson:2021ani}
A.~Kristensson, M.~Wilhelm, and C.~Zhang, {\it {Elliptic Double Box and
  Symbology Beyond Polylogarithms}},  {\em Phys. Rev. Lett.} {\bf 127} (2021),
  no.~25 251603, [\href{http://arxiv.org/abs/2106.14902}{{\tt
  arXiv:2106.14902}}].

\bibitem{Wilhelm:2022wow}
M.~Wilhelm and C.~Zhang, {\it {Symbology for elliptic multiple polylogarithms
  and the symbol prime}},  \href{http://arxiv.org/abs/2206.08378}{{\tt
  arXiv:2206.08378}}.

\bibitem{Morales:2022csr}
R.~Morales, A.~Spiering, M.~Wilhelm, Q.~Yang, and C.~Zhang, {\it {Bootstrapping
  elliptic Feynman integrals using Schubert analysis}},
  \href{http://arxiv.org/abs/2212.09762}{{\tt arXiv:2212.09762}}.

\bibitem{Bourjaily:2022bwx}
J.~L. Bourjaily et~al., {\it {Functions Beyond Multiple Polylogarithms for
  Precision Collider Physics}},  in {\em {2022 Snowmass Summer Study}}, 3,
  2022.
\newblock \href{http://arxiv.org/abs/2203.07088}{{\tt arXiv:2203.07088}}.

\bibitem{Bourjaily:2018ycu}
J.~L. Bourjaily, Y.-H. He, A.~J. Mcleod, M.~Von~Hippel, and M.~Wilhelm, {\it
  {Traintracks through Calabi-Yau Manifolds: Scattering Amplitudes beyond
  Elliptic Polylogarithms}},  {\em Phys. Rev. Lett.} {\bf 121} (2018), no.~7
  071603, [\href{http://arxiv.org/abs/1805.09326}{{\tt arXiv:1805.09326}}].

\bibitem{Bourjaily:2018yfy}
J.~L. Bourjaily, A.~J. McLeod, M.~von Hippel, and M.~Wilhelm, {\it {Bounded
  Collection of Feynman Integral Calabi-Yau Geometries}},  {\em Phys. Rev.
  Lett.} {\bf 122} (2019), no.~3 031601,
  [\href{http://arxiv.org/abs/1810.07689}{{\tt arXiv:1810.07689}}].

\bibitem{Bourjaily:2022tep}
J.~L. Bourjaily and N.~Kalyanapuram, {\it {The Stratification of Rigidity}},
  \href{http://arxiv.org/abs/2207.00596}{{\tt arXiv:2207.00596}}.

\bibitem{Bourjaily:2017bsb}
J.~L. Bourjaily, A.~J. McLeod, M.~Spradlin, M.~von Hippel, and M.~Wilhelm, {\it
  {Elliptic Double-Box Integrals: Massless Scattering Amplitudes beyond
  Polylogarithms}},  {\em Phys. Rev. Lett.} {\bf 120} (2018), no.~12 121603,
  [\href{http://arxiv.org/abs/1712.02785}{{\tt arXiv:1712.02785}}].

\bibitem{Paulos:2012nu}
M.~F. Paulos, M.~Spradlin, and A.~Volovich, {\it {Mellin Amplitudes for Dual
  Conformal Integrals}},  {\em JHEP} {\bf 08} (2012) 072,
  [\href{http://arxiv.org/abs/1203.6362}{{\tt arXiv:1203.6362}}].

\bibitem{Caron-Huot:2012awx}
S.~Caron-Huot and K.~J. Larsen, {\it {Uniqueness of two-loop master contours}},
   {\em JHEP} {\bf 10} (2012) 026, [\href{http://arxiv.org/abs/1205.0801}{{\tt
  arXiv:1205.0801}}].

\bibitem{Nandan:2013ip}
D.~Nandan, M.~F. Paulos, M.~Spradlin, and A.~Volovich, {\it {Star Integrals,
  Convolutions and Simplices}},  {\em JHEP} {\bf 05} (2013) 105,
  [\href{http://arxiv.org/abs/1301.2500}{{\tt arXiv:1301.2500}}].

\bibitem{Chicherin:2017bxc}
D.~Chicherin and E.~Sokatchev, {\it {Conformal anomaly of generalized form
  factors and finite loop integrals}},  {\em JHEP} {\bf 04} (2018) 082,
  [\href{http://arxiv.org/abs/1709.03511}{{\tt arXiv:1709.03511}}].

\bibitem{Bourjaily:2019hmc}
J.~L. Bourjaily, A.~J. McLeod, C.~Vergu, M.~Volk, M.~Von~Hippel, and
  M.~Wilhelm, {\it {Embedding Feynman Integral (Calabi-Yau) Geometries in
  Weighted Projective Space}},  {\em JHEP} {\bf 01} (2020) 078,
  [\href{http://arxiv.org/abs/1910.01534}{{\tt arXiv:1910.01534}}].

\bibitem{Vergu:2020uur}
C.~Vergu and M.~Volk, {\it {Traintrack Calabi-Yaus from Twistor Geometry}},
  {\em JHEP} {\bf 07} (2020) 160, [\href{http://arxiv.org/abs/2005.08771}{{\tt
  arXiv:2005.08771}}].

\bibitem{Bern:1994zx}
Z.~Bern, L.~J. Dixon, D.~C. Dunbar, and D.~A. Kosower, {\it {One loop n point
  gauge theory amplitudes, unitarity and collinear limits}},  {\em Nucl. Phys.
  B} {\bf 425} (1994) 217--260,
  [\href{http://arxiv.org/abs/hep-ph/9403226}{{\tt hep-ph/9403226}}].

\bibitem{Bern:1994cg}
Z.~Bern, L.~J. Dixon, D.~C. Dunbar, and D.~A. Kosower, {\it {Fusing gauge
  theory tree amplitudes into loop amplitudes}},  {\em Nucl. Phys. B} {\bf 435}
  (1995) 59--101, [\href{http://arxiv.org/abs/hep-ph/9409265}{{\tt
  hep-ph/9409265}}].

\bibitem{Cachazo:2008vp}
F.~Cachazo, {\it {Sharpening The Leading Singularity}},
  \href{http://arxiv.org/abs/0803.1988}{{\tt arXiv:0803.1988}}.

\bibitem{Frellesvig:2017aai}
H.~Frellesvig and C.~G. Papadopoulos, {\it {Cuts of Feynman Integrals in Baikov
  representation}},  {\em JHEP} {\bf 04} (2017) 083,
  [\href{http://arxiv.org/abs/1701.07356}{{\tt arXiv:1701.07356}}].

\bibitem{Bosma:2017ens}
J.~Bosma, M.~Sogaard, and Y.~Zhang, {\it {Maximal Cuts in Arbitrary
  Dimension}},  {\em JHEP} {\bf 08} (2017) 051,
  [\href{http://arxiv.org/abs/1704.04255}{{\tt arXiv:1704.04255}}].

\bibitem{Harley:2017qut}
M.~Harley, F.~Moriello, and R.~M. Schabinger, {\it {Baikov-Lee Representations
  Of Cut Feynman Integrals}},  {\em JHEP} {\bf 06} (2017) 049,
  [\href{http://arxiv.org/abs/1705.03478}{{\tt arXiv:1705.03478}}].

\bibitem{Dlapa:2021qsl}
C.~Dlapa, X.~Li, and Y.~Zhang, {\it {Leading singularities in Baikov
  representation and Feynman integrals with uniform transcendental weight}},
  {\em JHEP} {\bf 07} (2021) 227, [\href{http://arxiv.org/abs/2103.04638}{{\tt
  arXiv:2103.04638}}].

\bibitem{Abreu:2017ptx}
S.~Abreu, R.~Britto, C.~Duhr, and E.~Gardi, {\it {Cuts from residues: the
  one-loop case}},  {\em JHEP} {\bf 06} (2017) 114,
  [\href{http://arxiv.org/abs/1702.03163}{{\tt arXiv:1702.03163}}].

\bibitem{Flieger:2022xyq}
W.~Flieger and W.~J. Torres~Bobadilla, {\it {Landau and leading singularities
  in arbitrary space-time dimensions}},
  \href{http://arxiv.org/abs/2210.09872}{{\tt arXiv:2210.09872}}.

\bibitem{Arkani-Hamed:2010pyv}
N.~Arkani-Hamed, J.~L. Bourjaily, F.~Cachazo, and J.~Trnka, {\it {Local
  Integrals for Planar Scattering Amplitudes}},  {\em JHEP} {\bf 06} (2012)
  125, [\href{http://arxiv.org/abs/1012.6032}{{\tt arXiv:1012.6032}}].

\bibitem{Bourjaily:2020hjv}
J.~L. Bourjaily, N.~Kalyanapuram, C.~Langer, K.~Patatoukos, and M.~Spradlin,
  {\it {Elliptic, Yangian-Invariant \textquotedblleft{}Leading
  Singularity\textquotedblright{}}},  {\em Phys. Rev. Lett.} {\bf 126} (2021),
  no.~20 201601, [\href{http://arxiv.org/abs/2012.14438}{{\tt
  arXiv:2012.14438}}].

\bibitem{Frellesvig:2021vdl}
H.~Frellesvig, C.~Vergu, M.~Volk, and M.~von Hippel, {\it {Cuts and
  Isogenies}},  {\em JHEP} {\bf 05} (2021) 064,
  [\href{http://arxiv.org/abs/2102.02769}{{\tt arXiv:2102.02769}}].

\bibitem{Spradlin:2011wp}
M.~Spradlin and A.~Volovich, {\it {Symbols of One-Loop Integrals From Mixed
  Tate Motives}},  {\em JHEP} {\bf 11} (2011) 084,
  [\href{http://arxiv.org/abs/1105.2024}{{\tt arXiv:1105.2024}}].

\bibitem{Herrmann:2019upk}
E.~Herrmann and J.~Parra-Martinez, {\it {Logarithmic forms and differential
  equations for Feynman integrals}},  {\em JHEP} {\bf 02} (2020) 099,
  [\href{http://arxiv.org/abs/1909.04777}{{\tt arXiv:1909.04777}}].

\bibitem{Bourjaily:2019exo}
J.~L. Bourjaily, E.~Gardi, A.~J. McLeod, and C.~Vergu, {\it {All-mass $n$-gon
  integrals in $n$ dimensions}},  {\em JHEP} {\bf 08} (2020), no.~08 029,
  [\href{http://arxiv.org/abs/1912.11067}{{\tt arXiv:1912.11067}}].

\bibitem{Arkani-Hamed:2017ahv}
N.~Arkani-Hamed and E.~Y. Yuan, {\it {One-Loop Integrals from Spherical
  Projections of Planes and Quadrics}},
  \href{http://arxiv.org/abs/1712.09991}{{\tt arXiv:1712.09991}}.

\bibitem{Usyukina:1993ch}
N.~I. Usyukina and A.~I. Davydychev, {\it {Exact results for three and four
  point ladder diagrams with an arbitrary number of rungs}},  {\em Phys. Lett.
  B} {\bf 305} (1993) 136--143.

\bibitem{Broadhurst:2010ds}
D.~J. Broadhurst and A.~I. Davydychev, {\it {Exponential suppression with four
  legs and an infinity of loops}},  {\em Nucl. Phys. B Proc. Suppl.} {\bf
  205-206} (2010) 326--330, [\href{http://arxiv.org/abs/1007.0237}{{\tt
  arXiv:1007.0237}}].

\bibitem{Broadhurst:1993ib}
D.~J. Broadhurst, {\it {Summation of an infinite series of ladder diagrams}},
  {\em Phys. Lett. B} {\bf 307} (1993) 132--139.

\bibitem{Bourjaily:2019jrk}
J.~L. Bourjaily, F.~Dulat, and E.~Panzer, {\it {Manifestly Dual-Conformal Loop
  Integration}},  {\em Nucl. Phys. B} {\bf 942} (2019) 251--302,
  [\href{http://arxiv.org/abs/1901.02887}{{\tt arXiv:1901.02887}}].

\bibitem{Hodges:2009hk}
A.~Hodges, {\it {Eliminating spurious poles from gauge-theoretic amplitudes}},
  {\em JHEP} {\bf 05} (2013) 135, [\href{http://arxiv.org/abs/0905.1473}{{\tt
  arXiv:0905.1473}}].

\bibitem{Bouchard:2007ik}
V.~Bouchard, {\it {Lectures on complex geometry, Calabi-Yau manifolds and toric
  geometry}},  \href{http://arxiv.org/abs/hep-th/0702063}{{\tt
  hep-th/0702063}}.

\bibitem{Bonisch:2021yfw}
K.~B\"onisch, C.~Duhr, F.~Fischbach, A.~Klemm, and C.~Nega, {\it {Feynman
  integrals in dimensional regularization and extensions of Calabi-Yau
  motives}},  {\em JHEP} {\bf 09} (2022) 156,
  [\href{http://arxiv.org/abs/2108.05310}{{\tt arXiv:2108.05310}}].

\bibitem{Duhr:2022pch}
C.~Duhr, A.~Klemm, F.~Loebbert, C.~Nega, and F.~Porkert, {\it
  {Yangian-invariant fishnet integrals in 2 dimensions as volumes of Calabi-Yau
  varieties}},  \href{http://arxiv.org/abs/2209.05291}{{\tt arXiv:2209.05291}}.

\bibitem{Pogel:2022ken}
S.~P\"ogel, X.~Wang, and S.~Weinzierl, {\it {The $\varepsilon$-factorised
  differential equation for the four-loop equal-mass banana graph}},
  \href{http://arxiv.org/abs/2211.04292}{{\tt arXiv:2211.04292}}.

\bibitem{Chicherin:2020umh}
D.~Chicherin, J.~M. Henn, and G.~Papathanasiou, {\it {Cluster algebras for
  Feynman integrals}},  {\em Phys. Rev. Lett.} {\bf 126} (2021), no.~9 091603,
  [\href{http://arxiv.org/abs/2012.12285}{{\tt arXiv:2012.12285}}].

\bibitem{He:2021esx}
S.~He, Z.~Li, and Q.~Yang, {\it {Notes on cluster algebras and some all-loop
  Feynman integrals}},  {\em JHEP} {\bf 06} (2021) 119,
  [\href{http://arxiv.org/abs/2103.02796}{{\tt arXiv:2103.02796}}].

\bibitem{He:2021non}
S.~He, Z.~Li, and Q.~Yang, {\it {Truncated cluster algebras and Feynman
  integrals with algebraic letters}},  {\em JHEP} {\bf 12} (2021) 110,
  [\href{http://arxiv.org/abs/2106.09314}{{\tt arXiv:2106.09314}}]. [Erratum:
  JHEP 05, 075 (2022)].

\bibitem{He:2021eec}
S.~He, Z.~Li, and Q.~Yang, {\it {Kinematics, cluster algebras and Feynman
  integrals}},  \href{http://arxiv.org/abs/2112.11842}{{\tt arXiv:2112.11842}}.

\bibitem{Yang:2022gko}
Q.~Yang, {\it {Schubert problems, positivity and symbol letters}},  {\em JHEP}
  {\bf 08} (2022) 168, [\href{http://arxiv.org/abs/2203.16112}{{\tt
  arXiv:2203.16112}}].

\bibitem{He:2022ctv}
S.~He, Z.~Li, R.~Ma, Z.~Wu, Q.~Yang, and Y.~Zhang, {\it {A study of Feynman
  integrals with uniform transcendental weights and their symbology}},  {\em
  JHEP} {\bf 10} (2022) 165, [\href{http://arxiv.org/abs/2206.04609}{{\tt
  arXiv:2206.04609}}].

\bibitem{He:2022tph}
S.~He, J.~Liu, Y.~Tang, and Q.~Yang, {\it {The symbology of Feynman integrals
  from twistor geometries}},  \href{http://arxiv.org/abs/2207.13482}{{\tt
  arXiv:2207.13482}}.

\bibitem{Gaiotto:2011dt}
D.~Gaiotto, J.~Maldacena, A.~Sever, and P.~Vieira, {\it {Pulling the straps of
  polygons}},  {\em JHEP} {\bf 12} (2011) 011,
  [\href{http://arxiv.org/abs/1102.0062}{{\tt arXiv:1102.0062}}].

\bibitem{Caron-Huot:2016owq}
S.~Caron-Huot, L.~J. Dixon, A.~McLeod, and M.~von Hippel, {\it {Bootstrapping a
  Five-Loop Amplitude Using Steinmann Relations}},  {\em Phys. Rev. Lett.} {\bf
  117} (2016), no.~24 241601, [\href{http://arxiv.org/abs/1609.00669}{{\tt
  arXiv:1609.00669}}].

\bibitem{Caron-Huot:2020bkp}
S.~Caron-Huot, L.~J. Dixon, J.~M. Drummond, F.~Dulat, J.~Foster,
  O.~G\"urdo\u{g}an, M.~von Hippel, A.~J. McLeod, and G.~Papathanasiou, {\it
  {The Steinmann Cluster Bootstrap for $N$ = 4 Super Yang-Mills Amplitudes}},
  {\em PoS} {\bf CORFU2019} (2020) 003,
  [\href{http://arxiv.org/abs/2005.06735}{{\tt arXiv:2005.06735}}].

\bibitem{He:2021mme}
S.~He, Z.~Li, and Q.~Yang, {\it {Comments on all-loop constraints for
  scattering amplitudes and Feynman integrals}},  {\em JHEP} {\bf 01} (2022)
  073, [\href{http://arxiv.org/abs/2108.07959}{{\tt arXiv:2108.07959}}].
  [Erratum: JHEP 05, 076 (2022)].

\bibitem{He:2022ujv}
S.~He, Z.~Li, and C.~Zhang, {\it {A nice two-loop next-to-next-to-MHV amplitude
  in ${\cal N}=4$ super-Yang-Mills}},
  \href{http://arxiv.org/abs/2209.10856}{{\tt arXiv:2209.10856}}.

\bibitem{McLeod:2023qdf}
A.~McLeod, R.~Morales, M.~von Hippel, M.~Wilhelm, and C.~Zhang, {\it {An
  Infinite Family of Elliptic Ladder Integrals}},
  \href{http://arxiv.org/abs/2301.07965}{{\tt arXiv:2301.07965}}.

\bibitem{Bourjaily:2021vyj}
J.~L. Bourjaily, N.~Kalyanapuram, C.~Langer, and K.~Patatoukos, {\it
  {Prescriptive unitarity with elliptic leading singularities}},  {\em Phys.
  Rev. D} {\bf 104} (2021), no.~12 125009,
  [\href{http://arxiv.org/abs/2102.02210}{{\tt arXiv:2102.02210}}].

\end{thebibliography}\endgroup
\end{document}